\documentclass[AMA,LATO1COL]{WileyNJD-v2}

\articletype{Research article}

\pdfmapfile{+lato.map}

\usepackage{amssymb} 
\usepackage{textcomp}
\usepackage{booktabs}
\usepackage{amsmath}
\usepackage{hyperref}
\hypersetup{
	hidelinks = true,
}

\begin{document}

\title{A new wake-merging method for wind-farm power prediction in presence of heterogeneous background velocity fields}

\author{Luca Lanzilao}

\author{Johan Meyers}

\authormark{Lanzilao and Meyers}

\address{\orgdiv{Department of Mechanical Engineering}, \orgname{KU Leuven}, \orgaddress{\state{Leuven}, \country{Belgium}}}

\corres{{Luca Lanzilao, Department of Mechanical Engineering, KU Leuven, Leuven, Belgium.}
\email{luca.lanzilao@kuleuven.be}}

\abstract[Abstract]{Many wind farms are placed near coastal regions or in proximity of orographic obstacles. The meso-scale gradients that develop in these zones make wind farms operating in velocity fields that are rarely uniform. However, all existing wake-merging methods in engineering wind-farm wake models assume a homogeneous background velocity field in and around the farm, relying on a single wind-speed value usually measured several hundreds of meters upstream of the first row of turbines. In this study, we derive a new momentum-conserving wake-merging method capable of superimposing the waked flow on a heterogeneous background velocity field. We couple the proposed wake-merging method with four different wake models, i.e. the Gaussian, super-Gaussian, double-Gaussian and Ishihara model, and we test its performance against LES data, dual-Doppler radar measurements and SCADA data from the Horns Rev, London Array, and Westermost Rough farm. Next to this, as an additional point of reference, the standard Jensen model with quadratic superposition is also included. Results show that the new method performs similarly to linear superposition of velocity deficits in homogeneous conditions but it shows better performance when a spatially varying background velocity is used. The most accurate estimates are obtained when the wake-merging method is coupled with the double-Gaussian and Gaussian single-wake model. The Ishihara model also shows good agreements with observations. In contrast to this, the Jensen and super-Gaussian wake model underestimate the farm power output for all wind speeds, wind directions and wind farms considered in our analysis.}


\maketitle

\section{Introduction}\label{SecIntro} 

High cabling costs and shallow depths of water force the majority of offshore wind farms to be placed near coastal regions. Moreover, due to the current expansion of wind power capacity, the construction of onshore wind farms on non-flat complex terrains is common practice. The different surface roughness and temperature found over land and sea and the flow obstructions caused by orographic obstacles account for the development of strong velocity gradients \cite{Barthelmie2007,Vanderlaan2017,Durran1990}. Under such conditions, power and wake assessments performed with analytical wake models are challenging due to the heterogeneity of the background velocity field. In fact, the local value of wind speed and turbulence intensity measured several hundreds of meters upstream of the farm are usually the only flow characteristics used by existing wake models \cite{Lissaman1979,Katic1986,Voutsinas1990,Niayifar2016}. The goal of the current study is to develop a new momentum-conserving wake-merging method that accounts for a spatially varying background velocity and to further validate it against other superposition methods, numerical data and field measurements.

One of the first analytical single-wake model is the one proposed by Jensen \cite{Jensen1983}, which states that
\begin{equation}
\frac{\Delta\widehat{U}\bigl(x,r(\boldsymbol{x})\bigl)}{U_\mathrm{b}} =  \frac{1-\sqrt{1-C_\mathrm{T}}}{\bigl( 1+2\delta(x)/D \bigl)^2} \; \mathcal{H}\biggl(\frac{D}{2} + \delta(x) - r(\boldsymbol{x})\biggl)
\label{EqJensenModel}
\end{equation}
where $\boldsymbol{x} = (x,y,z)$ with $x$ the horizontal, $y$ the lateral and $z$ the vertical direction and $r(\boldsymbol{x}) = (y^2+z^2)^{1/2}$ denotes the radial direction. Moreover, $U_\mathrm{b}$ denotes the background velocity, $C_\mathrm{T}$ is the wind-turbine thrust coefficient, $\delta(x) = k^{\ast}x$ is the wake width with  $k^{\ast}$ the wake expansion coefficient, $D$ is the turbine rotor diameter, $\mathcal{H}$ denotes the Heaviside function and $\Delta \widehat{U}\bigl(x,r(\boldsymbol{x})\bigl)$ is the velocity deficit in the wake. Note that we are assuming the background velocity field to be uni-directional (i.e., only the horizontal component of the velocity field is non-zero). This model is derived applying only the mass conservation law in a control volume located downwind of the turbine \cite{Bastankhan2014}. Despite the lack of conservation of momentum, the Jensen model is one of the most used in both academic studies \cite{Gonzalez2010,Pena2016} and commercial software \cite{Windfarmer2009,Truepower2010}. Later, Frandsen et al \cite{Frandsen2006} proposed a new wake model based on both mass and momentum conservation laws applied on a control volume located around the turbine. Both the Jensen and Frandsen model assume a top-hat shape for the velocity deficit in the wake. This assumption leads to an underestimation of $\Delta \widehat{U}(x,r(\boldsymbol{x}))$ at the center of the wake and to an overestimation of the velocity deficit at the edges. Since the power depends on the cube of the wind speed, the top-hat shape assumption can lead to large errors when used for predicting farms energy output \cite{Niayifar2016,Bastankhan2014}.

A different approach consists in using self-similarity in the wake \cite{Pope2000} so that the velocity deficit is expressed as 
\begin{equation}
\frac{\Delta\widehat{U}\bigl(x,r(\boldsymbol{x})\bigl)}{U_\mathrm{b}} =  C(x) f\bigl(r(\boldsymbol{x}),\delta(x)\bigl)
\label{EqSelfSimilarity}
\end{equation}
where $C(x)$ denotes the maximum normalized velocity deficit at each downwind location, $\delta(x)$ is the wake width and $f\bigl(r(\boldsymbol{x}),\delta(x)\bigl)$ describes the shape of the velocity profile. While the shape function $f\bigl(r(\boldsymbol{x}),\delta(x)\bigl)$ is prescribed, the maximum normalized velocity deficit $C(x)$ is derived from the mass and momentum conservation laws \cite{Bastankhan2014}. Table \ref{Table1} shows three different $C(x)$ relations derived with distinct shape functions. The Gaussian wake model \cite{Bastankhan2014} assumes a Gaussian-like velocity deficit profile. It has proven to outperform both the Jensen and Frandsen model \cite{Bastankhan2014} and is nowadays widely used in wake assessments and layout optimization problems \cite{Xiaoxia2016,Parada2017}. However, this model does not conserve momentum in the near-wake region. To improve this deficiency, the super-Gaussian wake model uses a top-hat shape function in the near-wake region which evolves to a Gaussian shape in the far wake \cite{Shapiro2019,Blondel2020}. A different approach to improve the Gaussian wake model in the near-wake region is offered by the double-Gaussian wake model, which includes the nacelle effects using two Gaussian functions which are symmetric with respect to the wake center. Results have shown that this model outperforms the Gaussian model in the near-wake region, but it overestimates the velocity deficit far downstream the turbine \cite{Schreiber2020}. Other types of models prescribe the shape function and derive the maximum velocity deficit $C(x)$ by fitting of numerical and experimental data. The model proposed by Ishihara and Qian \cite{Ishihara2018} is an example, where $C(x) = (a+bx/D+p)^{-2}$ with $a$, $b$, and $p$ model parameters obtained from fitting of data.

\begin{table}[]
	\centering
	\caption{Shape function $\widehat{W}(x,r(\boldsymbol{x})) = C(x) f\bigl(r(\boldsymbol{x}),\delta(x)\bigl)$ and maximum normalized velocity deficit of three momentum-conserving wake models. Note that for $x\leq0$, $C(x)=0$.}
	\begin{tabular}{clclc}
		\textbf{Model}  &  & $f(r,\delta)$                                                                           &  & $C(x)~~~~(x>0)$                                                                     \\ \hline
		\addlinespace[0.2cm]
		Gaussian        &  & $\displaystyle \exp \biggl(-\frac{r^2}{2 \delta^2} \biggl)$                                                                     &  & $\displaystyle 1 - \sqrt{1-\frac{C_\mathrm{T}}{8(\delta/D)^2}}$                                \\
		\addlinespace[0.2cm]
		Super-Gaussian  &  & $\displaystyle \exp \biggl( -\frac{r^n}{2 \delta^2} D^{2-n} \biggl)$                                                                     &  & $\displaystyle 2^{2/n-1} - \sqrt{2^{4/n-2} - \frac{nC_\mathrm{T}}{16\Gamma(2/n) \bigl(\delta/D \bigl)^{4/n}}} $ \\
		\addlinespace[0.2cm]
		Double-Gaussian &  & $\displaystyle \frac{1}{2}\biggl[  \exp \biggl( -\frac{(r+r_0)^2}{2 \delta^2} \biggl) + \exp \biggl(-\frac{(r-r_0)^2}{2 \delta^2} \biggl) \biggl]$ &  & $\displaystyle \frac{M-\sqrt{M^2 - 1/2 N C_\mathrm{T} D^2}}{2N} $                               \\ \addlinespace[0.2cm] \hline
		
	\end{tabular}
	\label{Table1}
\end{table}

\begin{figure}[t!]
	\centering
	\includegraphics[width=0.45\textwidth]{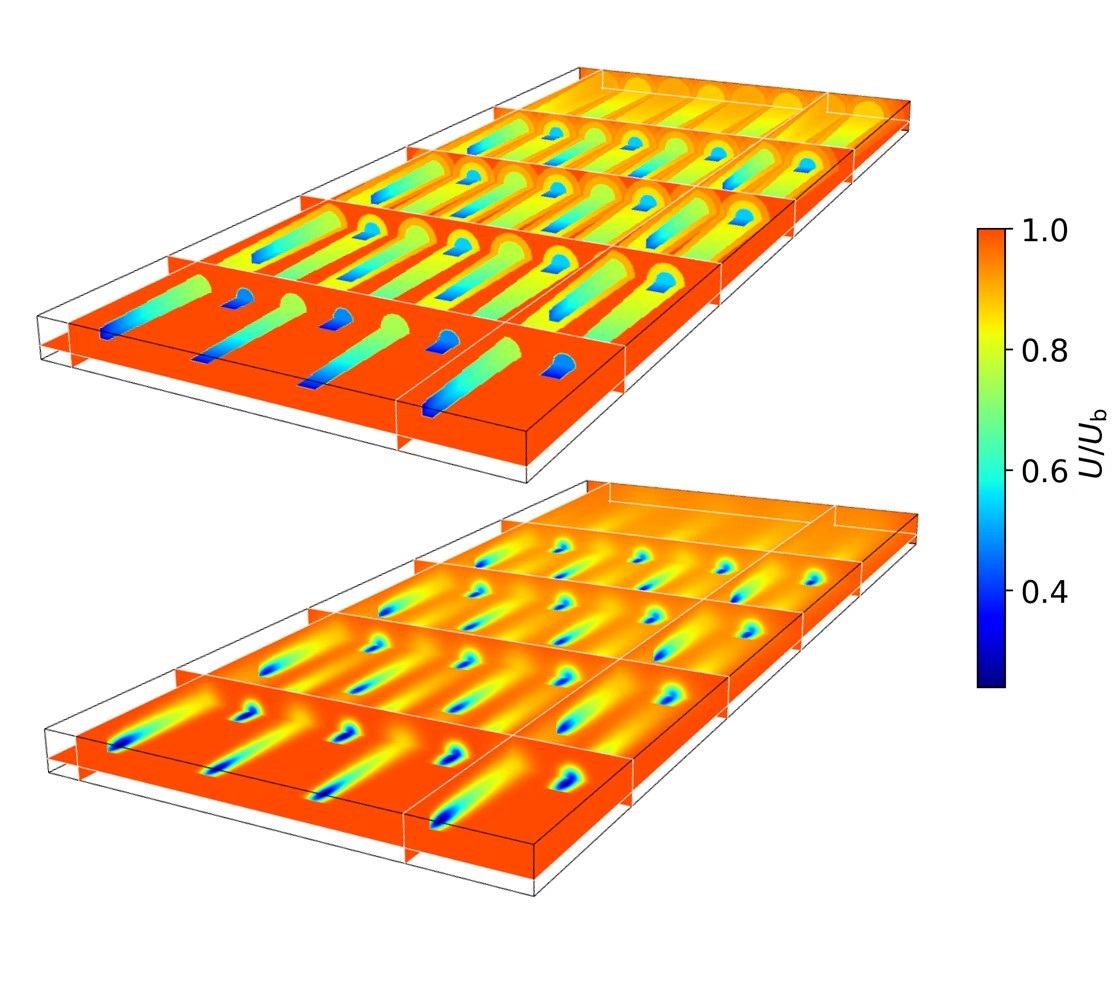}%
	\caption{Normalized velocity field through a staggered farm with 32 turbines with rotor diameter $D=$ 154 m, thrust coefficient $C_\mathrm{T}=$ 0.85, ambient turbulence intensity $\mathrm{TI}_\mathrm{b}=$ 12\% and background velocity $U_\mathrm{b}=$ 10 m/s. The dimensionless streamwise and spanwise spacing are $s_x=s_y=$ 6. (top) Jensen model combined with superposition of energy deficit and (bottom) Gaussian model coupled with linear superposition of velocity deficit.}
	\label{FigJensenNiayifar}
\end{figure}

The wake models mentioned above are used for wake assessments of single wind turbines. However, to estimate the power output of large farms, we have to deal with multiple wakes. Hence, superposition methods are applied to account for wake overlapping. Many wake-merging methods exist in literature \cite{Machefaux2015,Zong2020} but, to date, the most used are the two following
\begin{align}
U(\boldsymbol{x}) &=  U_\mathrm{b} - \sum_{k=1}^{N_t} u_k W_k(\boldsymbol{x}) \label{Niayifar}\\
U(\boldsymbol{x}) &=  U_\mathrm{b} - \sqrt{\sum_{k=1}^{N_t} \bigl( u_k W_k(\boldsymbol{x}) \bigl)^2} \label{Voutsinas}
\end{align}
where $U(\boldsymbol{x})$ is the horizontal component of the waked flow while $u_k$ denotes the wind velocity averaged over the rotor disk of turbine $k$. To define the wake function $W_k(\boldsymbol{x})$, we define a new coordinate system $\boldsymbol{\xi}_k(\boldsymbol{x}) = \bigl(\zeta_k(\boldsymbol{x}),\chi_k(\boldsymbol{x}),\eta_k(\boldsymbol{x}) \bigl)$ per turbine which has origin in correspondence of the location of turbine $k$, which we denote with $\boldsymbol{x}_k = (x_k,y_k,z_{h,k})$, with $z_{h,k}$ the turbine hub-height. Hence, $\zeta_k(\boldsymbol{x}) = x - x_k$, $\chi_k(\boldsymbol{x}) = y - y_k$, $\eta_k(\boldsymbol{x}) = z - z_{h,k}$ and $W_k(\boldsymbol{x}) = \widehat{W}\bigl(\zeta_k(\boldsymbol{x}),r(\boldsymbol{\xi}_k(\boldsymbol{x}))\bigl)$ with
\begin{equation}
\widehat{W}\bigl(\zeta_k,r(\boldsymbol{\xi}_k)\bigl) = C\bigl(\zeta_k\bigl) f\bigl(r(\boldsymbol{\xi}_k),\delta(\zeta_k)\bigl).
\label{EqWhat}
\end{equation}
Note that the maximum normalized velocity deficit and the wake width assume non-zero values only for $\zeta_k(\boldsymbol{x})>0$. Moreover, the formulation above holds under the assumption of uni-directional flow. Since the product $u_k W_k(\boldsymbol{x})$ corresponds to a velocity deficit (see Eq.~\ref{EqSelfSimilarity}), Eq.~\ref{Niayifar} proposed by Niayifar and Porté-Agel\cite{Niayifar2016} evaluates the velocity field within the farm using a linear superposition of velocity deficits while Eq.~\ref{Voutsinas} proposed by Voutsinas et al\cite{Voutsinas1990} uses a superposition of energy deficits. Figure \ref{FigJensenNiayifar} displays a three-dimensional representation of the velocity field within a staggered farm with dimensionless streamwise and spanwise spacing of $s_x = s_y=$ 6. The Jensen model combined with Eq.~\ref{Voutsinas} is used in Fig. \ref{FigJensenNiayifar} (top) while the Gaussian model coupled with Eq.~\ref{Niayifar} is adopted in Fig. \ref{FigJensenNiayifar} (bottom). The plots reveal that a more smooth flow field through the farm is obtained with a Gaussian shape function. A more extensive comparison between the two models is performed in Section \ref{Result}.

Despite the numerous wake-merging methods developed over the years, none of them accounts for a possible spatially varying background velocity field, i.e $\boldsymbol{U}_\mathrm{b}(\boldsymbol{x}) = \bigl( U_\mathrm{b}(\boldsymbol{x}),V_\mathrm{b}(\boldsymbol{x}) \bigl)$. Instead, a global value of wind speed is used which is not always trivial to define due to possible heterogeneity in the velocity field. The only wake model that accounts for varying background velocity is the one proposed by Shamsoddin and Porté-Agel\cite{Shamsoddin2018}. However, this model deals with the wake of single turbines and it has never been extended to a farm level. In the current manuscript, we aim to fill the aforementioned gap by developing a new momentum-conserving wake-merging method that accounts for spatial heterogeneity in the background flow field. The manuscript is structured as follows. Firstly, Section~\ref{NewModel} introduces the new wake-merging method and shows some of its features. Next, Section~\ref{Setup} describes the single-wake model setup and details the simulation cases. Thereafter, we compare the performance of the new wake-merging method against LES data, dual-Doppler radar measurements and SCADA data in Section~\ref{Result}. Finally, conclusions and suggestions for future research are drawn in Section~\ref{conclusion}.

\section{New wake-merging method}\label{NewModel}
In Section~\ref{SecUniFLow}, we derive the new wake-merging method for a uni-directional flow which changes in magnitude. Thereafter, in Section~\ref{SecMultiFlow}, we generalize the model considering a background velocity field which changes in direction over the wind farm area. Finally, some example cases which highlight the features of the new model are illustrated in Section~\ref{SecExampleCase}. 

\subsection{Uni-directional flow} \label{SecUniFLow}
Consider a single row of $N_t$ turbines that are aligned with the flow. The turbines operate in a heterogeneous uni-directional background velocity field which we denote with $U_\mathrm{b}(\boldsymbol{x})$. To compute the waked flow through the farm, we use a recursive formula. The velocity field at the first iteration is given by the background wind speed, which is a model input. Hence, $U_0(\boldsymbol{x}) = U_\mathrm{b}(\boldsymbol{x})$. Using self-similarity in the wake of the most upwind turbine (i.e., Eq~\ref{EqSelfSimilarity}), we can write the velocity field $U_1(\boldsymbol{x})$ as
\begin{equation}
\frac{U_0(\boldsymbol{x}) - U_1(\boldsymbol{x})}{U_0(\boldsymbol{x})} = W_1(\boldsymbol{x})
\qquad
\Longrightarrow
\qquad
U_1(\boldsymbol{x}) = U_0(\boldsymbol{x}) \bigl[1 - W_1(\boldsymbol{x}) \bigl].
\label{EqSelfSim1D}
\end{equation}
More generally, given the flow field generated by the background field and the first $k-1$ turbines, the effect of the next turbine ($k$) is also expressed using self-similarity of the wake (of turbine $k$), so that
\begin{equation}
\frac{U_{k-1}(\boldsymbol{x}) - U_k(\boldsymbol{x})}{U_{k-1}(\boldsymbol{x})} = W_k(\boldsymbol{x})
\qquad
\Longrightarrow
\qquad
U_k(\boldsymbol{x}) = U_{k-1}(\boldsymbol{x}) \bigl[1 - W_k(\boldsymbol{x}) \bigl], 
\label{EqUniFlowRecursive}
\end{equation}
providing the general recursion formula for any value of $k$. Finally the full flow field in the farm (background and all wakes) corresponds with $U(\boldsymbol{x})= U_{N_t}(\boldsymbol{x})$. For the current case of uni-directional flow, the recursion  relation (Eq.~\ref{EqUniFlowRecursive}) can be explicitly elaborated into 
\begin{equation}
U(\boldsymbol{x}) = U_\mathrm{b}(\boldsymbol{x}) \prod_{k=1}^{N_t} \bigl[1-W_k(\boldsymbol{x}) \bigl], \qquad \mbox{with} \qquad W_k(\boldsymbol{x}) = \widehat{W}\bigl(\zeta_k(\boldsymbol{x}),r(\boldsymbol{\xi}_k(\boldsymbol{x}))\bigl).
\label{NewWakeModel}
\end{equation}
Note that Eq.~\ref{NewWakeModel} holds for an arbitrary wind-farm layout. In fact, the wake function $W_k(\boldsymbol{x})$ is defined in such a way that $W_k(\boldsymbol{x})=0$ for any location $\boldsymbol{x}$ upstream of turbine $k$.

\subsection{Multi-directional flow}\label{SecMultiFlow}
We now generalize the ideas discussed above for multi-directional flow. To this end, we consider an arbitrary farm layout now immersed in a heterogeneous background velocity field which changes direction and magnitude throughout the farm. Hence, $\boldsymbol{U}_\mathrm{b}(\boldsymbol{x}) = \bigl(U_\mathrm{b}(\boldsymbol{x}),V_\mathrm{b}(\boldsymbol{x}) \bigl)$, and $\theta_\mathrm{b}(x,y)=\arctan\bigl(V_\mathrm{b}(\boldsymbol{x})/U_\mathrm{b}(\boldsymbol{x})\bigl)$ defines the orientation of the background flow. For sake of simplicity, we keep the turbines aligned with the flow (i.e, no yaw misalignment). We also assume that the vertical velocity component of the background velocity field is zero, and that the background flow does not change direction along the $z$-axis (meaning that the ratio $V_\mathrm{b}(\boldsymbol{x})/U_\mathrm{b}(\boldsymbol{x})$ is constant with height, i.e., that effects of wind veer are negligible). 

First of all, we now presume that the wakes in the farm are carried by the background flow $\boldsymbol{U}_\mathrm{b}(\boldsymbol{x})$. Thus, the coordinate system  $\boldsymbol{\xi}_k(\boldsymbol{x}) = \bigl(\zeta_k(\boldsymbol{x}),\chi_k(\boldsymbol{x}),\eta_k(\boldsymbol{x}) \bigl)$ used to express the wake function of turbine $k$ is now oriented along the streamlines of the background flow field, with origin  the turbine location (i.e., $\boldsymbol{\xi}_k(\boldsymbol{x}_k)=\boldsymbol{0}$). Hence, 
\begin{align}
\zeta_k(\boldsymbol{x}) &= \int_{x_k}^x \cos{\bigl(\theta_\mathrm{b}(\bar{x},y)\bigl)} d\bar{x} + \int_{y_k}^y \sin{\bigl(\theta_\mathrm{b}(x,\bar{y})\bigl)} d\bar{y} \label{eq:coord1}\\
\chi_k(\boldsymbol{x}) &= -\int_{x_k}^x \sin{\bigl(\theta_\mathrm{b}(\bar{x},y)\bigl)} d\bar{x} + \int_{y_k}^y \cos{\bigl(\theta_\mathrm{b}(x,\bar{y})\bigl)} d\bar{y} \label{eq:coord2}\\
\eta_k(\boldsymbol{x}) &= z - z_{h,k}. \label{eq:coord3}
\end{align}

Next, we construct the recursion formula. To this end, we first order the turbines from most upstream to most downstream, so that for any $l>k$, the wake of turbine $l$ does not interfere with turbine $k$. The starting term of the recursion is then given again by $\boldsymbol{U}_0(\boldsymbol{x}) = \boldsymbol{U}_\mathrm{b}(\boldsymbol{x})$. The recursion formula now becomes
\begin{equation}
\boldsymbol{U}_{k}(\boldsymbol{x}) = \bigl(\boldsymbol{U}_{k-1}(\boldsymbol{x}) \cdot \boldsymbol{e}_{\perp,k}\bigl) \bigl( 1 - W_k(\boldsymbol{x}) \bigl) \boldsymbol{e}_{\perp,k} + \bigl(\boldsymbol{U}_{k-1}(\boldsymbol{x}) \cdot \boldsymbol{e}_{\parallel,k}\bigl) \boldsymbol{e}_{\parallel,k}, \qquad \mbox{for }\;\; k=1,\dots,N_t
\label{EqRecursive}
\end{equation}
and the total farm flow field corresponds to $\boldsymbol{U}(\boldsymbol{x}) = \boldsymbol{U}_{N_t}(\boldsymbol{x})$. The wake deficit function $W_k(\boldsymbol{x}) = \widehat{W}\bigl(\zeta_k(\boldsymbol{x}), r(\boldsymbol{\xi}_k(\boldsymbol{x}))\bigl)$ is defined as in Eq. \ref{EqWhat}. Moreover, $\boldsymbol{e}_{\perp,k} = \bigl(\cos{\theta_k}, \sin{\theta_k} \bigl)$ and $\boldsymbol{e}_{\parallel,k} = \bigl( -\sin{\theta_k},\cos{\theta_k} \bigl)$ denote the unit vectors perpendicular and parallel to the rotor plane of turbine~$k$. Thus, Eq.~\ref{EqRecursive} simply expresses that the wake deficit $\bigl(1-W_k(\boldsymbol{x})\bigl)$ is oriented in the axial direction of turbine $k$, while no velocity change occurs in the parallel direction. This axial velocity deficit is then transported along the streamlines of the background flow as expressed by the coordinate system $\boldsymbol{\xi}_k(\boldsymbol{x})$.

Finally, the angle $\theta_k$ corresponds to the orientation of turbine $k$, and is defined as 
\begin{equation}
\theta_k\triangleq \arctan\bigl(V(\boldsymbol{x}_k)/U(\boldsymbol{x}_k)\bigl) = \arctan\bigl(V_{k-1}(\boldsymbol{x}_k)/U_{k-1}(\boldsymbol{x}_k)\bigl), \label{eq:turborient}
\end{equation} 
where the second equality follows from the ordering of the turbines. Further, $\theta_1 = \theta_\mathrm{b}(x_1,y_1)$, but $\theta_k$ is in general different from $\theta_\mathrm{b}(x_k,y_k)$. Note that we have used the simplifying assumption that the coordinate system  $\boldsymbol{\xi}_k(\boldsymbol{x})$ (Eqs.~\ref{eq:coord1}--\ref{eq:coord3}) is oriented along the background flow. This can be further generalized so that $\boldsymbol{\xi}_k$ is oriented along the streamlines of the full flow $\boldsymbol{U}(\boldsymbol{x})$. However, this would require an iterative procedure, i.e., starting with orientation along the background flow as an initial guess, and then iteratively updating the streamlines using Eqs.~\ref{eq:coord1}--\ref{eq:turborient} (e.g. based on a Newton method or a fixed-point iteration method). This is however, not further considered here.

Equation \ref{EqRecursive} can also be further generalized to include effects of yaw misalignment. In such case, the orientation of turbine $k$ would be expressed as $\widetilde{\theta}_k = \theta_k + \gamma_k$ with $\gamma_k$ the yaw angle of turbine $k$ and the wake deficit function would need to be extended to include yaw effects (see Bastankhah and Porté-Agel\cite{Bastankhah2016} and Qian and Ishihara\cite{Qian2018} for examples). Moreover, Eq.~\ref{EqRecursive} would have to account for a velocity deficit also in the parallel direction.  However, the effects of yaw misalignment are not included in the current manuscript. 

In the discussion above, the wake function $\widehat{W}(x,r(\boldsymbol{x}))$ can be selected from Table \ref{Table1}. However, the derivation of $C(x)$ assumes that the background velocity field upstream of the turbine is uniform in the stream and cross-stream directions \cite{Bastankhan2014,Anderson2011}. Instead, the background velocity perceived by the turbine $k$ is non-uniform due to heterogeneity in the background velocity field and preceding wakes (see, e.g., Eq.~\ref{EqUniFlowRecursive}). Hence, the maximum normalized velocity deficit proposed by Shamsoddin and Porté-Agel \cite{Shamsoddin2018} should be adopted. However, the assumptions of slow varying background velocity in the horizontal directions (i.e., the x- and y-derivative of both components of $\boldsymbol{U}_\mathrm{b}(\boldsymbol{x})$ are negligible) and large enough turbine spacings allow us to obtain the same parameters $C(x)$ reported in Table \ref{Table1}. In other words, Eq.~\ref{EqRecursive} is mass and momentum-conserving only if a slow varying background velocity is used (which is usually the case for velocity changes induced by coastal gradients -- see Barthelmie et al\cite{Barthelmie2007} and Section~\ref{Westermost}) and if preceding wakes generate smooth velocity changes. 

Finally, the inflow velocity of the turbine $k$ for the computation of its power (or thrust force) is obtained by computing the average velocity over the disk. To this end, we use a quadrature rule with $N_q=$ 16 points, spread over the turbine rotor disk. The quadrature-point coordinates on the rotor disk of the turbine $k$ are denoted with $\boldsymbol{x}_{k,q}$ and are chosen following the quadrature rule proposed by Holoborodko \cite{Holoborodko2011} with uniform weighting factor $\omega_q = 1/N_q$. Hence
\begin{equation}
u_k = \sum_{q=1}^{N_q} \omega_q S(\boldsymbol{x}_{k,q})
\label{EqTurbineInflowVel}
\end{equation}
where $S(\boldsymbol{x}) = \Vert \boldsymbol{U}(\boldsymbol{x}) \Vert_2$. We refer to Allaerts and Meyers\cite{Allaerts2019} for more details about the quadrature rule. The turbine power is then obtained from $P_k = P_\mathrm{c}^k(u_k)$ where $P_\mathrm{c}^k$ denotes the power curve of turbine $k$.

\subsection{Example cases} \label{SecExampleCase}
To visually illustrate the new wake-merging methodology, we visualize a couple of simple examples here. A validation of the method against data is performed later in Section~\ref{Result}. 

\begin{figure}[t!]
	\centering
	\includegraphics[width=1.\textwidth]{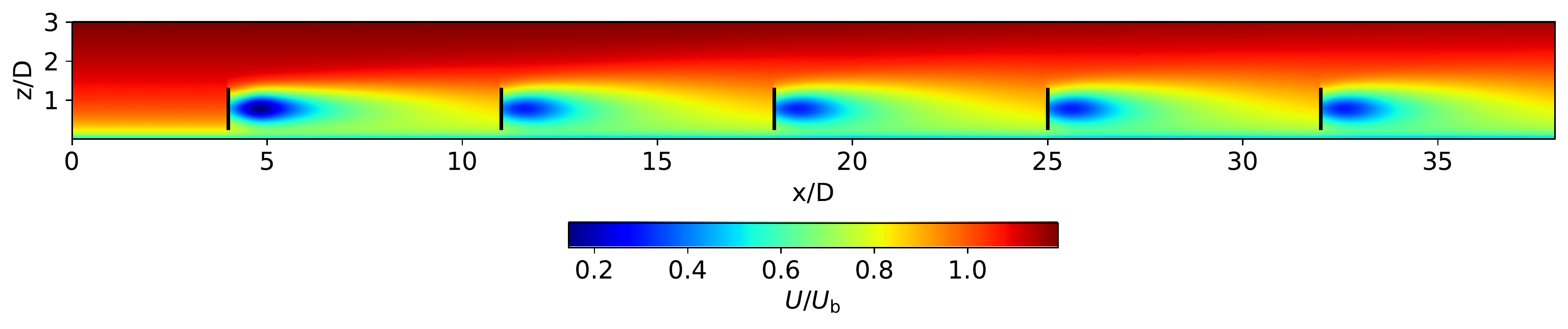}%
	\caption{Normalized velocity field at a vertical plane normal to the wind turbine at zero span for a single row of 5 turbines with dimensionless streamwise spacing of $s_x=$ 7. The velocity field is computed with the new wake-merging method coupled with the Gaussian wake model. The velocity $U_\mathrm{b}$ is homogenous in the x- and y-direction. The black lines denote the wind-turbine rotor locations.}
	\label{FigSideViewNewModel}
\end{figure}

The background velocity $\boldsymbol{U}_\mathrm{b}(\boldsymbol{x})$ is a model input. For instance, $\boldsymbol{U}_\mathrm{b}(\boldsymbol{x})$ could be the three-dimensional free-stream velocity taken in the vicinity of an operating farm (see Nygaard and Newcombe\cite{Nygaard2018} and Djath et al\cite{Djath2018}) or, ideally, the velocity field which would manifest assuming that the farm would not operate. If measurements data are not available, the background velocity can, e.g., be modelled as a horizontally homogeneous field with a logarithmic profile\cite{Tennekes1973} in the vertical direction. Figure~\ref{FigSideViewNewModel} shows an example of a row of 5 turbines with streamwise spacing of $s_x=$ 7 immersed in such a velocity field. We assume uni-directional flow (i.e., $\theta_\mathrm{b}=0^\circ$) with $U_\mathrm{b} = U_\mathrm{b}(z)$, hence we fix the friction velocity and the surface roughness to 0.58 m/s and 0.1 m, respectively. The wake function is the one proposed by Bastankhan and Porté-Agel\cite{Bastankhan2014} with $C_\mathrm{T}=$ 0.85. The ambient turbulence intensity is $\mathrm{TI}_\mathrm{b}=$ 12\% and the added turbulence is evaluated adopting the model proposed by Niayifar and Porté-Agel\cite{Niayifar2016}. It is worth to notice the formation of an internal boundary layer \cite{Wu2013,Allaerts2017} which grows along the streamwise direction. Moreover, the velocity recovery in the wake of the turbines can be appreciated. Note that ground-wake interactions are neglected throughout the whole manuscript, hence the model would show an axisymmetric velocity distribution if no vertical shear is considered. Moreover, the Gaussian wake model does not conserve momentum in the near-wake region, i.e. when $C_\mathrm{T} > D^2/8\delta^2$ (see Table \ref{Table1}). To represent the wake deficits in this region, the thrust coefficients are written as an error function of the streamwise coordinate, similarly to Zong and Porté-Agel\cite{Zong2020}.

\begin{figure}[t!]
	\centering
	\begin{minipage}{\textwidth}
		\centering
		\includegraphics[width=0.4\textwidth]{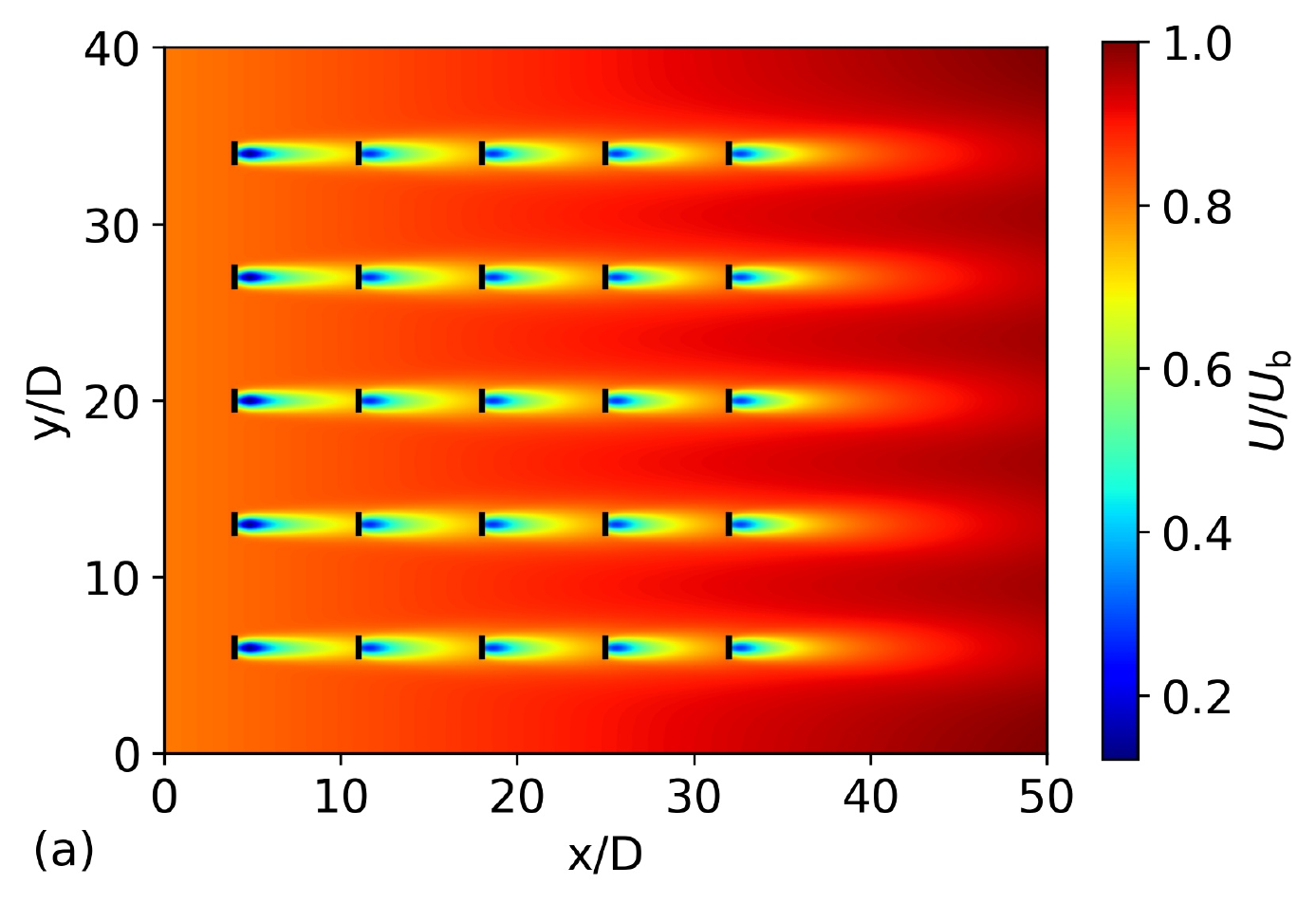}%
		\qquad\quad
		\hspace{0.003\textwidth}
		\includegraphics[width=0.4\textwidth]{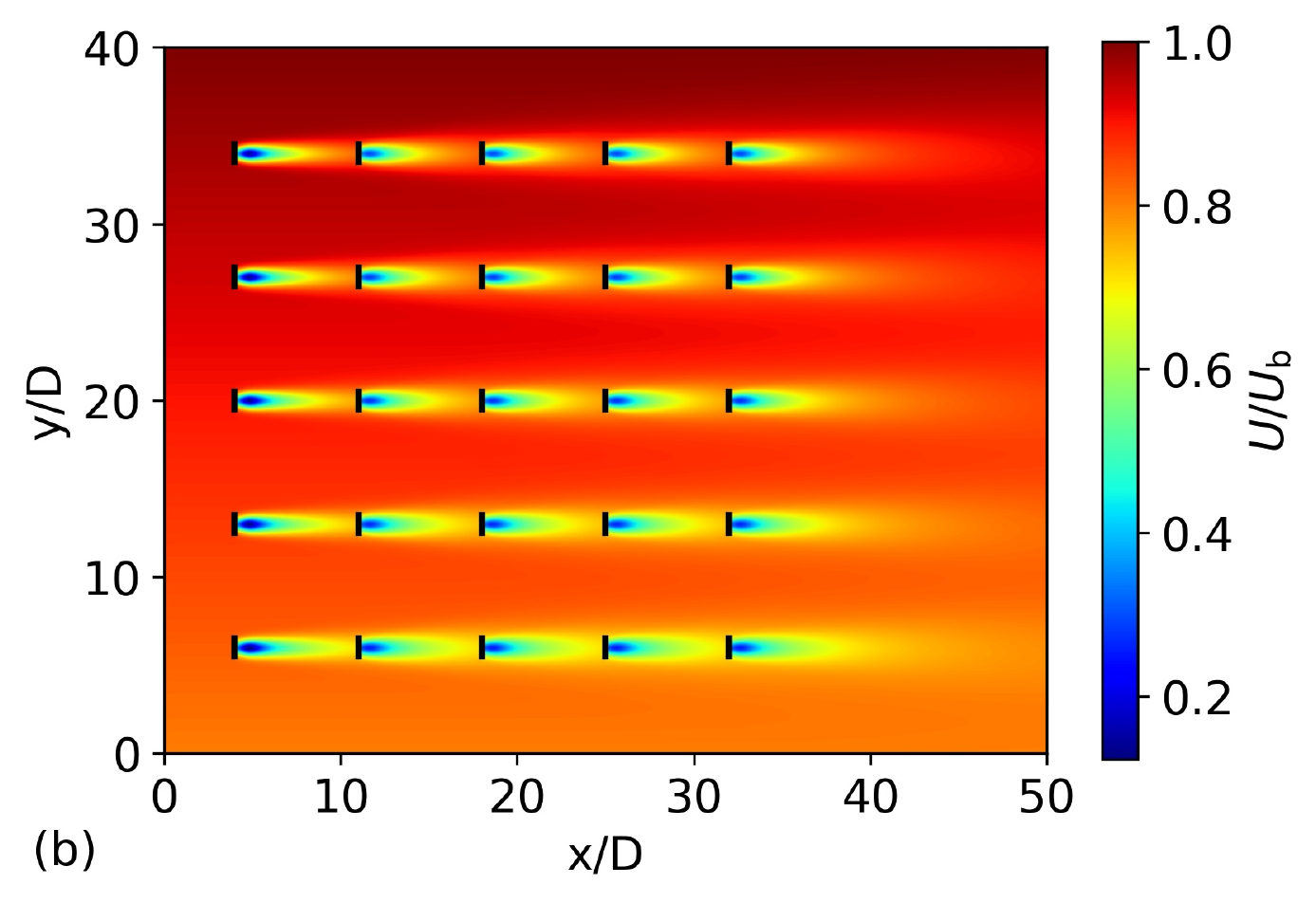}%
		\qquad\quad
		\vspace{0.003\textwidth}
		\includegraphics[width=0.4\textwidth]{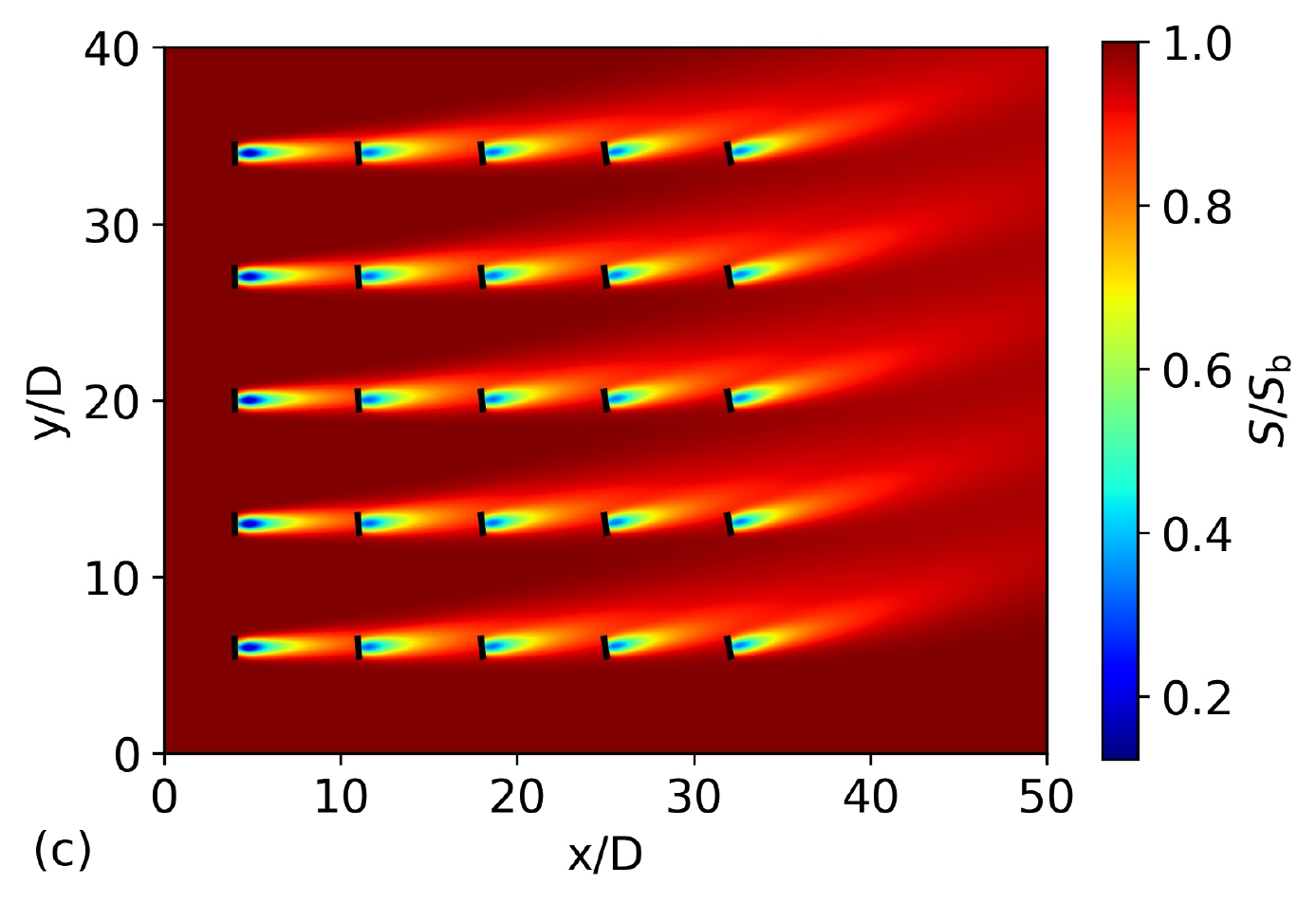}%
		\caption{Normalized velocity field at a horizontal plane at hub height computed with the new wake-merging method coupled with the Gaussian wake model. (a,b) Abrupt change of background velocity magnitude along the streamwise and spanwise directions, respectively. (c) Abrupt change of background velocity direction over the wind farm. The farm contains 25 turbines with dimensionless streamwise and spanwise spacing of $s_x = s_y=$ 7, respectively. The black lines denote the wind-turbine rotor locations.}
		\label{FigNewModelDeviation}
	\end{minipage}
\end{figure}

The same model setup is used in Fig. \ref{FigNewModelDeviation}(a,b), but now a spatially varying background velocity $U_\mathrm{b}(x,y)$ is adopted. More specifically, Fig. \ref{FigNewModelDeviation}(a,b) illustrates the turbines' wake behaviour in presence of an abrupt change in background velocity magnitude along the streamwise and spanwise direction, respectively (i.e., the background wind speed increases linearly along the streamwise (spanwise) direction with a difference of 2 m/s between $x/D=$ 0 ($y/D=$ 0) and $x/D=$ 50 ($y/D=$ 40)). Thereafter, Fig.~\ref{FigNewModelDeviation}(c) shows the flow through a farm which is subject to an abrupt change in wind direction. In this specific case, we assume $S_\mathrm{b} = \Vert \boldsymbol{U}_\mathrm{b}(\boldsymbol{x}) \Vert_2 =$ 10 m/s and we vary the wind direction with $\theta_\mathrm{b}(x) = 0.35 x/D$, so that there is a 10° change in background wind direction between the first and the last column of turbines.  

In the current section, we have shown that the proposed wake-merging method properly accounts for preceding wakes (i.e., Eq.~\ref{EqSelfSimilarity} is satisfied for every turbine). Moreover, the new method does not rely on a single velocity value taken upstream of the farm but instead it accounts for heterogeneous background velocity fields. Despite the added features, the computation of the velocity through a farm can be done within seconds, making the model suitable for optimization studies and annual energy yield assessments. A more extensive model validation is performed in Section \ref{Result}, where we compare the model's performance against numerical data and field measurements in homogeneous and heterogeneous conditions. 

\section{Cases description and single-wake model setup}\label{Setup}
We compare nine wind-farm models: the Gaussian\cite{Bastankhan2014} (G), super-Gaussian \cite{Blondel2020} (SG), double-Gaussian \cite{Schreiber2020} (DG) and Ishihara\cite{Ishihara2018}(I) single-wake model coupled with linear superposition of velocity deficits (Lin -- Eq.~\ref{Niayifar}) and with our new wake-merging method (New -- Eq.~\ref{EqRecursive}). As an additional reference, we also show results from the Jensen model \cite{Jensen1983} combined with quadratic superposition (Eq.~\ref{Voutsinas}). In the remainder of this section, we will briefly describe the setup of the aforementioned single-wake models. For sake of simplicity, we assume uni-directional flow condition and that the stand-alone turbine is located in position $x_1=y_1=0$, so that no horizontal axes transformation is needed.

The Jensen model, which we only use as a standard point of reference in the current work, evaluates the normalized velocity deficit using Eq.~\ref{EqJensenModel}, which depends on the wake expansion coefficient $k^{\ast}$, the only free-parameter. Jensen \cite{Jensen1983} considered $k^{\ast}=$ 0.1, however we assume $k^{\ast}=$ 0.04 as it is normally used for offshore wind-farm studies \cite{Nygaard2018,Cleve2009,Barthelmie2010b,Nygaard2014}. Note that the wake decay parameter does not depend on the turbulence intensity level.

\begin{figure}[t!]
	\centering
	\includegraphics[width=1.\textwidth]{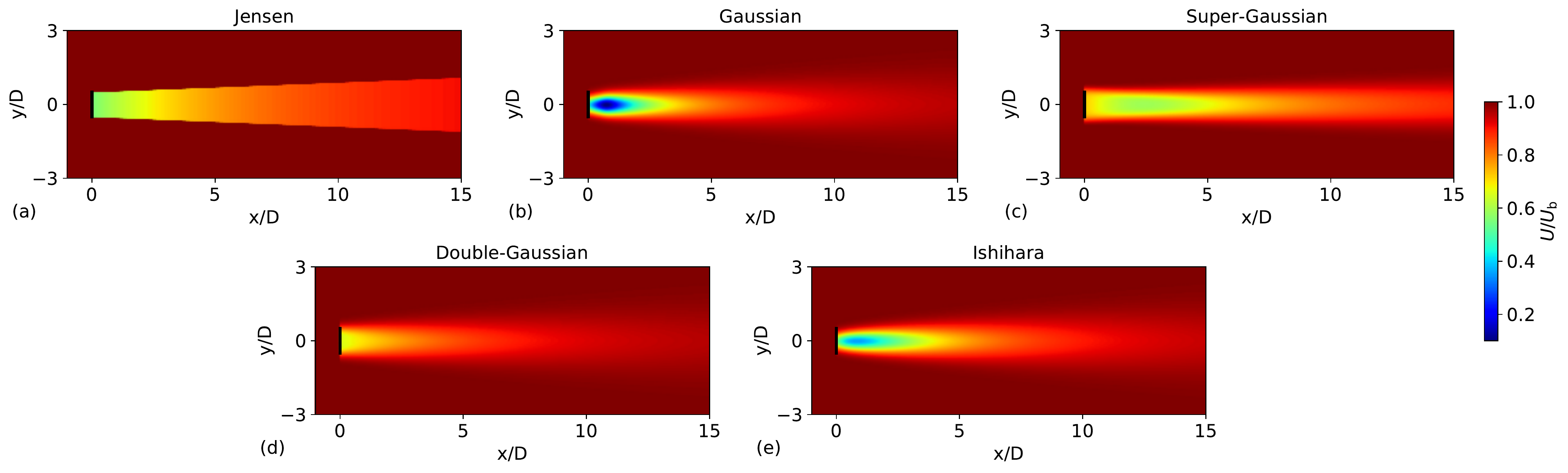}%
	\caption{Normalized velocity field at a horizontal plane at hub height computed with the (a) Jensen, (b) Gaussian, (c) super-Gaussian, (d) double-Gaussian and (e) Ishihara single-wake model. The black lines denote the wind-turbine rotor locations.}
	\label{FigSingleWakeModels}
\end{figure}

The Gaussian wake model used in our analysis is the one proposed by Bastankhan and Porté-Agel\cite{Bastankhan2014}. The shape function and parameter $C(x)$ are reported in Table \ref{Table1}. Similarly to Jensen\cite{Jensen1983}, a linear expansion of the wake width is assumed, i.e.
\begin{equation}
\delta(x) = k^{\ast}x+\epsilon D
\label{EqWakeWidthBast}
\end{equation}
where $\epsilon=0.2 \sqrt{\beta}$ is the limit value of $\delta(x)/D$ for $x$ approaching zero, with
\begin{equation}
\beta = \frac{1}{2} \frac{1+\sqrt{1-C_\mathrm{T}}}{\sqrt{1-C_\mathrm{T}}}.
\end{equation}	
Contrary to the Jensen model, the wake expansion coefficient is expressed as a function of the turbulence intensity evaluated at turbine location following the empirical equation proposed by Niayifar and Porté-Agel\cite{Niayifar2016}, that is $k^{\ast}=$ 0.3837$ \mathrm{TI} +$ 0.003678. Note that this empirical equation provides good estimates of $k^{\ast}$ for turbulence intensity values in the range of 0.065 $<\mathrm{TI}<$0.15.

The choice of a super-Gaussian shape function implies a $n$-dependent maximum normalized velocity deficit $C(x)$ (see Table \ref{Table1}). For high values of $n$, the shape function resembles a top-hat filter, while for low values the function evolves towards a Gaussian shape. If $n=$ 2, the model is identical to the Gaussian one. Blondel and Cathelain\cite{Blondel2020} proposed the following expression for the order $n(x)$
\begin{equation}
n(x) = a_{\!f} e^{b_{\!f}x/D} + c_{\!f}.
\end{equation}
Regarding the wake width, the following linear form is used
\begin{equation}
\delta(x) = \bigl[a_s \mathrm{TI} + b_s \bigl] x + c_s \sqrt{\beta} D.
\end{equation}
Following Blondel and Cathelain\cite{Blondel2020}, we use $a_{\!f}= 3.11$, $b_{\!f}=-0.68$, $c_{\!f}=2.41$, $a_s=0.17$, $b_s=0.005$ and $c_s=0.2$. These parameters are derived from fitting of numerical data and field measurements. Note that with the current parameters selection, the super-Gaussian wake model does not converge to the Gaussian one in the far-wake region. We refer to the original publication for more details \cite{Blondel2020}.

In an attempt to improve the Gaussian wake model in the near-wake region, a double-Gaussian shape function is proposed by Schreiber et al\cite{Schreiber2020}. In this model, the parameter $C(x)$ (see Table \ref{Table1}) depends upon two analytically derived functions
\begin{align}
M(x) &= 2 \delta^2(x) \exp \biggl(- \frac{r_0^2}{2 \delta^2(x)} \biggl)+ \sqrt{2 \pi} r_0 \delta(x) \text{ erf} \biggl( \frac{r_0}{\sqrt{2}\delta(x)}\biggl) \label{EqDGM} \\
N(x) &= \delta^2(x) \exp \biggl(- \frac{r_0^2}{\delta^2(x)} \biggl)+ \frac{\sqrt{\pi}}{2} r_0 \delta(x) \text{ erf} \biggl( \frac{r_0}{\delta(x)}\biggl) \label{EqDGN}
\end{align}
where $r_0 = Dk_r/2$ represents the radial position of the Gaussian extrema. Following Schreiber et al\cite{Schreiber2020}, we set $k_r=$ 0.535 in the current manuscript. The wake width expression used by Schreiber et al\cite{Schreiber2020} reads as
\begin{equation}
\delta(x) = k^*(x-s_0) + \alpha D
\label{EqWakeWidthDG}
\end{equation}
with $s_0=4.55D$ the stream tube outlet position and $\alpha = 0.23D$ the wake expansion at location $s_0$. However, we have noticed that this wake width parametrization leads to a strong overestimation of the velocity deficits in the far-wake region. This causes unrealistic power losses predictions when the model is coupled with superposition methods (not shown). To avoid the issue, we use in Eqs.~\ref{EqDGM} and \ref{EqDGN} the wake width expression proposed by Bastankhan and Porté-Agel\cite{Bastankhan2014} (i.e., Eq.~\ref{EqWakeWidthBast}) instead of Eq.~\ref{EqWakeWidthDG}.

The only model considered in our analysis which is not analytically derived is the one proposed by Ishihara and Qian\cite{Ishihara2018}. This model assumes a Gaussian shape function and it uses a parameter $C(x)$ defined as
\begin{equation}
C(x) = \frac{1}{\bigl( a + bx/D + c (1+x/D)^{-2}\bigl)^2}
\end{equation}
with 
\begin{align}
a &= 0.93 C_\mathrm{T}^{-0.75} \mathrm{TI}^{0.17},\\
b &= 0.42 C_\mathrm{T}^{0.6} \mathrm{TI}^{0.2},\\
c &= 0.15 C_\mathrm{T}^{-0.25} \mathrm{TI}^{-0.7}.
\end{align}
The expressions of these parameters are derived from fitting of numerical data and wind tunnel experiments (see Ishihara and Qian\cite{Ishihara2018} for more information). 

Velocity fields at a horizontal plane at hub height for a stand-alone wind turbine computed with the five single-wake models described above are illustrated in Fig.~\ref{FigSingleWakeModels}. The thrust set-point of the turbine is $C_\mathrm{T}=$ 0.7 with $D=$ 154 m. The ambient turbulence intensity is $\mathrm{TI}_\mathrm{b}=$ 12\% and a homogeneous uni-directional background velocity field is used, hence $U_\mathrm{b}=$ 10 m/s. It is worth to notice that the ratio $U/U_\mathrm{b}$ predicted by the Jensen and super-Gaussian wake model is close to 0.9 at 14$D$ downstream of the turbine (see Fig.~\ref{FigSingleWakeModels}(a) and Fig.~\ref{FigSingleWakeModels}(c), respectively), which is low if compared with results of the other single-wake models. Moreover, the Gaussian, double-Gaussian and Ishihara model predict a similar wake profile far downstream the turbine and differ only in the near-wake region (i.e., when $x/D<$ 4). Note that the Gaussian model does not conserve momentum when $C_\mathrm{T} < D^2/8 \delta^2$ (which corresponds to $x/D<$ 3/2 in the current example). The thrust coefficient is written as an error function of the streamwise coordinate to simulate the wake deficits in this region (see Zong and Porté-Agel\cite{Zong2020}).

In the remainder of the manuscript, we denote with Lin--G, Lin--SG, Lin--DG, Lin--I, New--G, New--SG, New--DG, New--I the eight wind-farm models used in our analysis, where the first word specifies the superposition method, the hyphen has to be read as ``coupled with'' and the second word identifies the single-wake model. Moreover, the Jensen model coupled with quadratic superposition is simply denoted with ``Jensen''. All wind-farm models with a turbulence-dependent wake decay coefficient use the method proposed by Niayifar and Porté-Agel\cite{Niayifar2016} (which is inspired to the work of Frandsen and Th{\o}gersen\cite{Frandsen1999}) for computing the added turbulence through the farm.

\section{Results and discussion}\label{Result}
The aim of this section is to compare the performance of the linear superposition method introduced by Niayifar and Porté-Agel\cite{Niayifar2016} with the new wake-merging method. As an add additional point of reference, we also include the Jensen model in the analysis. The models power predictions are compared against LES and SCADA data from the Horns Rev and London Array wind farm in Section \ref{HornsRev} and \ref{LondonArray}, respectively. Thereafter, a velocity field, reconstructed from dual-Doppler radar measurements taken at the Westermost Rough farm, is used as a reference in Section \ref{Westermost}. The farms characteristics and types of data used are reported in Table \ref{Table2}. Note that we assume a vertically-homogeneous background flow with constant wind direction throughout the whole section. More insights about the differences between the new and the linear superposition method are provided in Appendix~\ref{AppNewNia}.

\begin{table}[]
	\centering
	\caption{Farms characteristics and types of data used.}
	\begin{tabular}{ccccccc}
		\textbf{Wind farm} & \textbf{Capacity {[}MW{]}} & \textbf{N° of turbines} & \textbf{Hub height {[}m{]}} & \textbf{Rotor diameter {[}m{]}} & \textbf{Turbine}    & \textbf{Data}   \\ \hline
		\addlinespace[0.1cm]
		Horns Rev          & 160                        & 80                          & 70                          & 80                              & Vestas V-80  & LES, SCADA         \\
		\addlinespace[0.1cm]
		London Array       & 630                        & 175                         & 84.5                        & 120                             & SWT-120-3.6 & SCADA \\
		\addlinespace[0.1cm]
		Westermost Rough   & 210                        & 35                          & 106                         & 154                             & SWT-154-6.0   &  Dual-Doppler radar        \\
		\addlinespace[0.1cm] \hline
	\end{tabular}
	\label{Table2}
\end{table}

\subsection{Horns Rev} \label{HornsRev}
\begin{figure}[t!]
	\centering
	\begin{minipage}{\textwidth}
		\centering
		\includegraphics[width=0.313\textwidth]{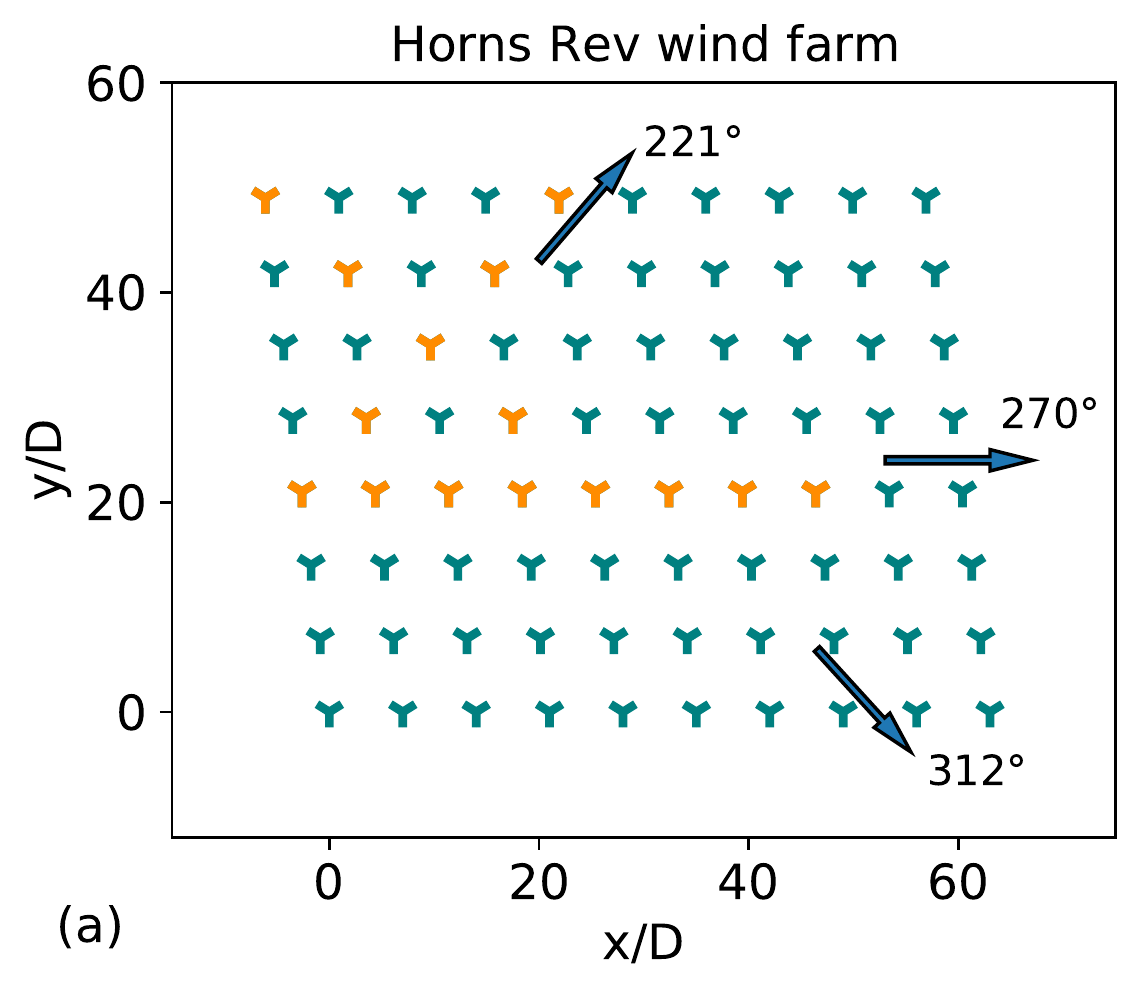}%
		\qquad\quad
		\hspace{0.003\textwidth}
		\includegraphics[width=0.442\textwidth]{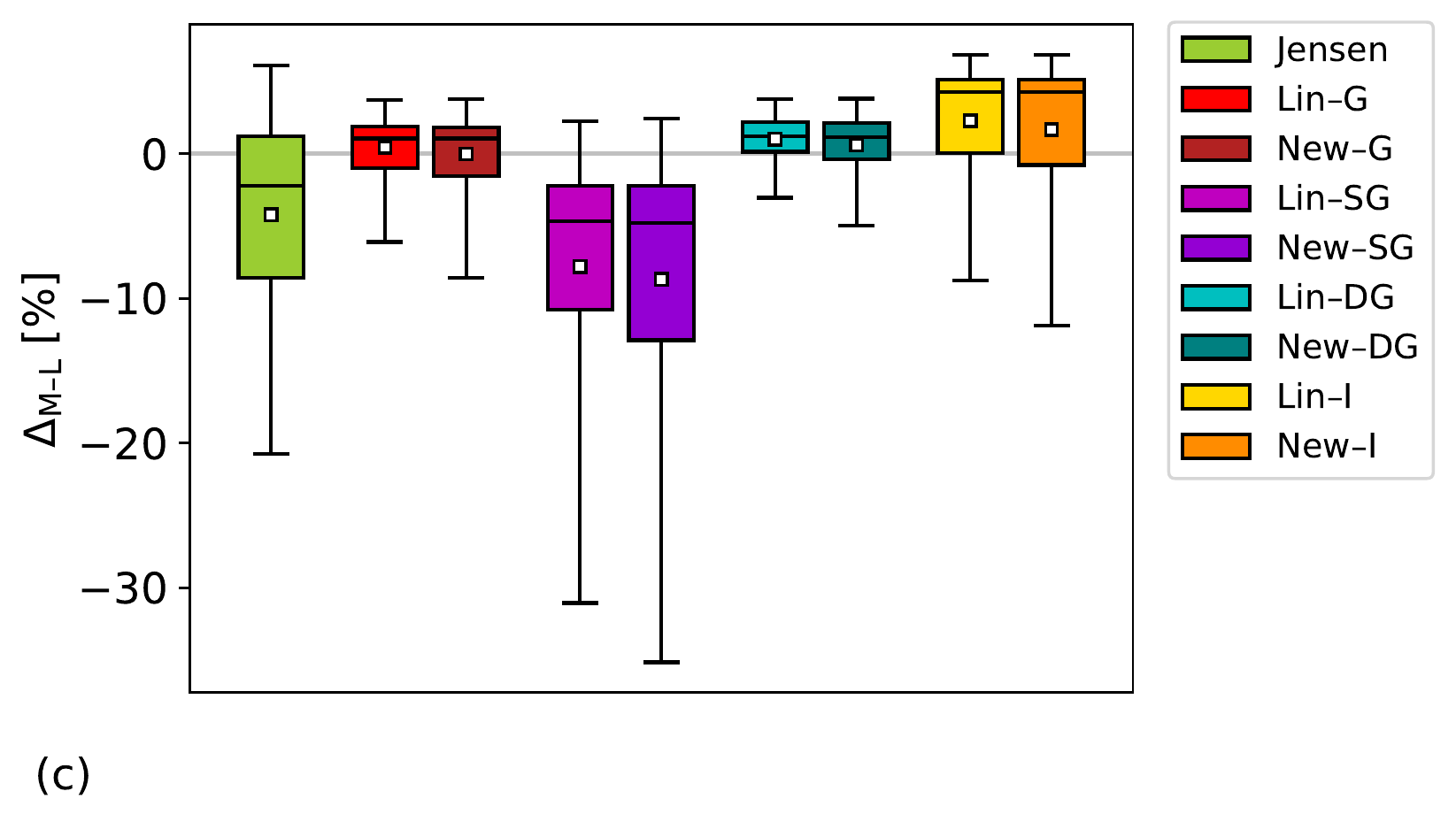}%
		\vspace{0.015\textwidth}
		\includegraphics[width=0.9\textwidth]{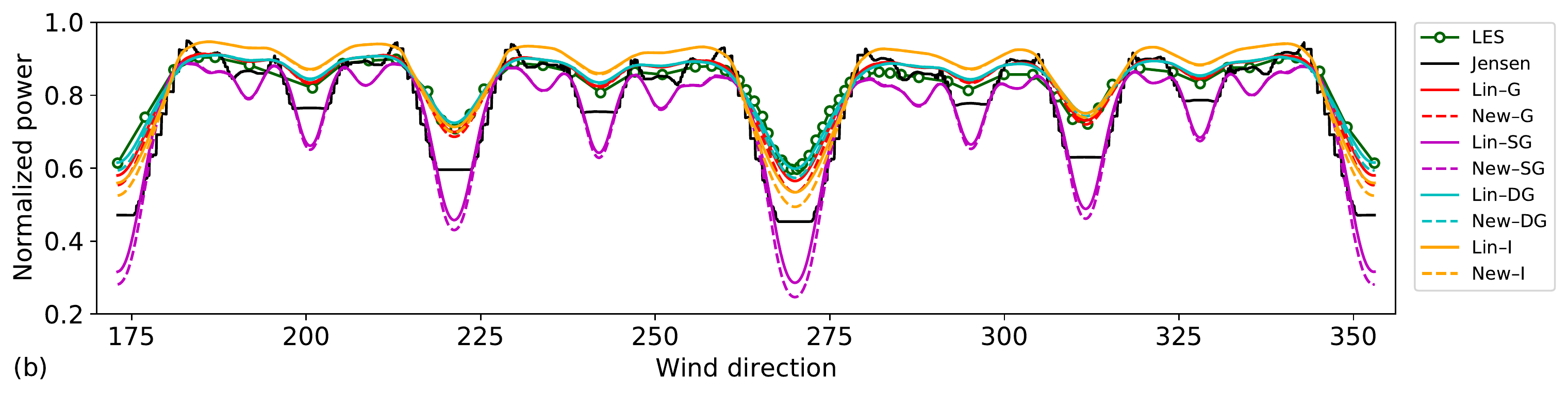}%
		\caption{(a) Layout of the Horns Rev farm with distances normalized with the rotor diameter. SCADA data are only available for turbines marked in orange. (b) Horns Rev power output normalized with the wake-less power for changing wind directions computed with LES and  nine analytical wind-farm models . (c) Distribution of the difference in farm power output between models predictions and LES data for changing wind directions normalized with the wake-less power. The box length represents the interquartile range. The horizontal black line and the white square denote the median and the mean, respectively. The whiskers include outliers, hence the caps represent the maximum and minimum $\Delta_\mathrm{M\text{--}L}$ value.}
		\label{FigHornsRevLES}
	\end{minipage}
\end{figure}

The Horns Rev was the first offshore wind farm in the North Sea, located 15 km off the westernmost point of Denmark. Eighty Vestas V-80 are laid out as an oblique rectangle with a dimensionless streamwise and spanwise spacing of $s_x=s_y=$ 7. The farm layout is shown in Fig. \ref{FigHornsRevLES}(a) while the farm characteristics are detailed in Table \ref{Table2}. The Vestas V-80 thrust and power curve are provided by Niayifar and Porté-Agel\cite{Niayifar2016}. The average free-stream velocity is assumed to be constant, hence $U_\mathrm{b}=$ 8 m/s, with ambient turbulence intensity of 7.7\%. The SCADA data \cite{Barthelmie2009} and LES results \cite{Niayifar2016} refer to the same inflow conditions.

Figure \ref{FigHornsRevLES}(b) illustrates the Horns Rev power output for different wind directions (from 173° to 353°) computed with a LES (see Wu and Porté-Agel\cite{Wu2013} and Niayifar and Porté-Agel \cite{Niayifar2016} for more details) and with the nine farm models discussed in this manuscript. Note that the power is normalized with the wake-less power (i.e., the power that the farm would extract if all turbines would operate in non-waked conditions). Overall, the new wake-merging method performs similarly to the Niayifar one, with a slightly higher power underestimation occurring for low turbine spacing values (i.e., for wind direction of 221°, 270° and 312°). To better summarize the wind-farm models performance, Fig. \ref{FigHornsRevLES}(c) displays the distribution of the difference between models ($P_\mathrm{Mod}^\mathrm{wf}$) and LES ($P_\mathrm{LES}$) farm power output for changing wind directions normalized with the wake-less power, which is computed as
\begin{equation}
\Delta_\mathrm{M\text{--}L} = \frac{P_\mathrm{Mod}^\mathrm{wf} - P_\mathrm{LES}}{N_t P_1}
\end{equation}
where $N_t$ is the total number of wind turbines and $P_1$= 0.6 MW is the power extracted by the freestream turbine. To put things into perspective, a $\Delta_\mathrm{M\text{--}L}$ of 10\% corresponds to a difference in farm power output between model and LES data of 4.8 MW (3\% of the farm rated power). Note that positive values of $\Delta_\mathrm{M\text{--}L}$ denote an overestimation of farm power output. Both the Gaussian and double-Gaussian wake model show very accurate power prediction, with biases close to zero and a maximum power underestimation of about 8\%, which occurs at 270°. On the other hand, the Jensen and super-Gaussian wake model show negative biases and wider interquartile ranges, and perform poorly when the farm is exposed to wind directions for which there is a small streamwise distance between turbines (see Fig. \ref{FigHornsRevLES}(b)). The bad performance obtained with the super-Gaussian model could be attributed to a sub-optimal choice of the tuning parameters. In fact, the choice of $c_{\!f}=$ 2.41 leads to higher velocity deficits in the far-wake region with respect to a Gaussian wake model, as shown in Fig.~\ref{FigSingleWakeModels}(c). Finally, the Ishihara model, which is the only non-analytically derived model, shows fairly good agreement with LES results. Note that the Jensen model is the only one that uses a top-hat shape function, which explains the staircase (and less realistic) power behaviour observed in Fig. \ref{FigHornsRevLES}(b).

\begin{figure}[t!]
	\centering
	\begin{minipage}{\textwidth}
		\centering
		\vspace{0.005\textwidth}
		\includegraphics[width=0.95\textwidth]{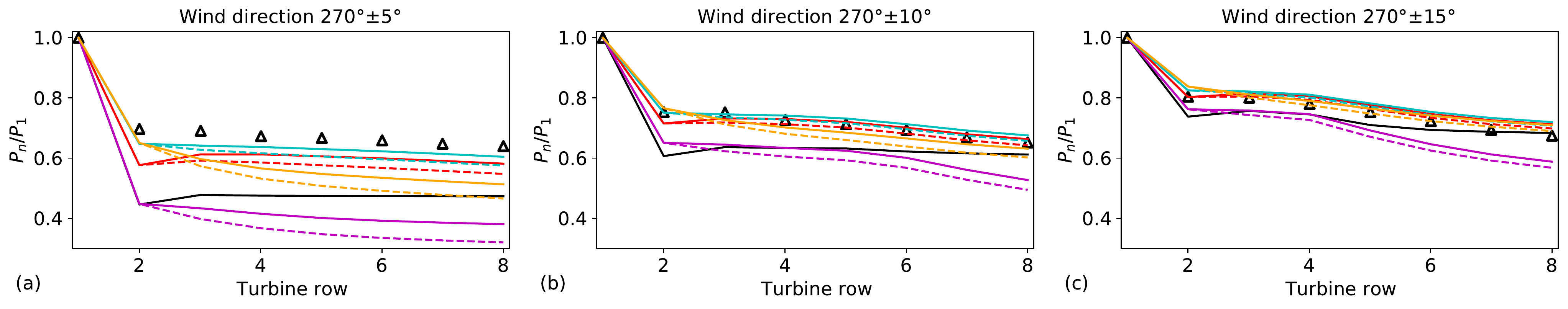}%
		\vspace{0.005\textwidth}
		\includegraphics[width=0.95\textwidth]{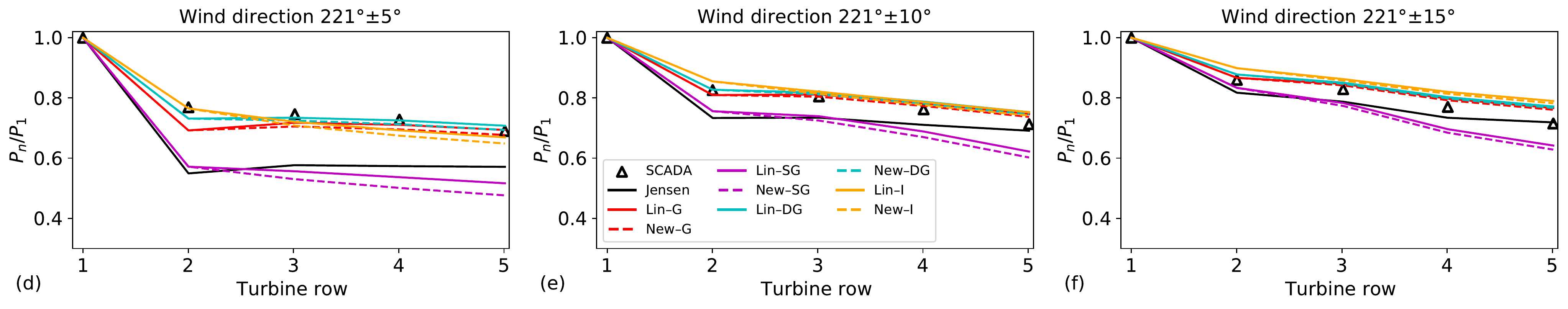}%
		\vspace{0.005\textwidth}
		\includegraphics[width=0.95\textwidth]{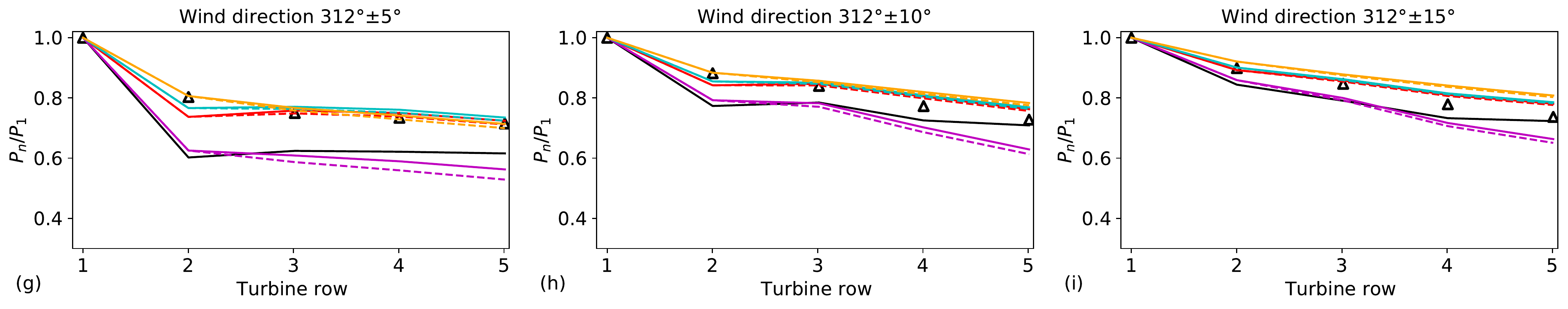}%
		\caption{Power ratios averaged over a wind sector of (left column) $\pm$5°, (middle column) $\pm$10° and (right column) $\pm$15° centered on a mean wind direction of (top row) 270°, (middle row) 221° and (bottom row) 312°.}
		\label{FigHornsRevPowerRow}
	\end{minipage}
\end{figure}

\begin{figure}[t!]
	\centering
	\includegraphics[width=0.98\textwidth]{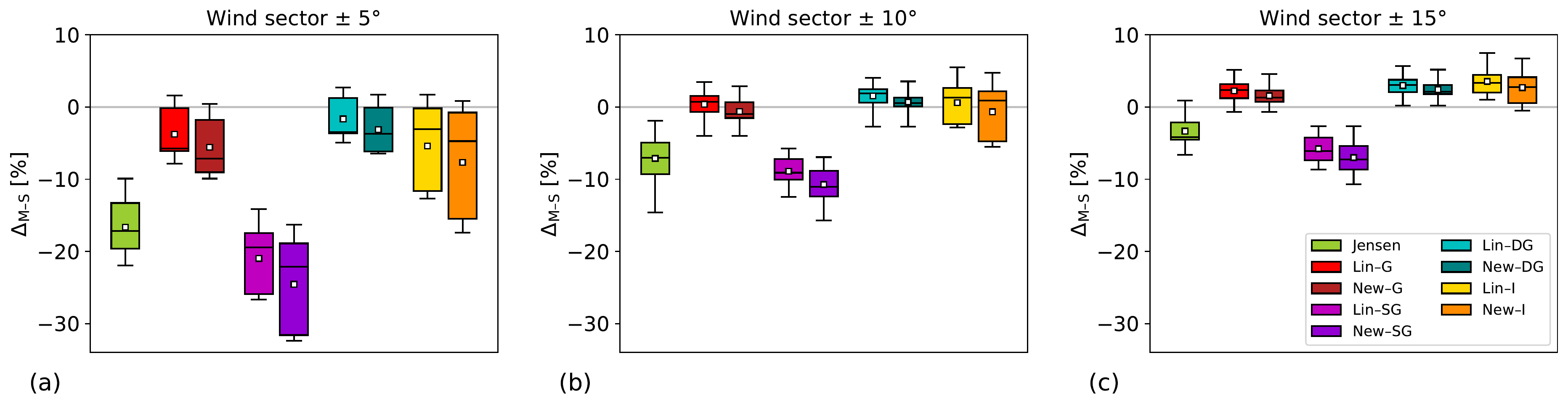}%
	\caption{Distribution of the difference in turbine power output between models predictions and SCADA data for a wind sector of (a) $\pm$5°, (b) $\pm$10° and (c) $\pm$ 15° centered on three mean wind directions (221°, 270° and 312°). The difference in turbine power output is normalized with the freestream power. The box length represents the interquartile range. The horizontal black line and the white square denote the median and the mean, respectively. The whiskers include outliers, hence the caps represent the maximum and minimum $\Delta_\mathrm{M\text{--}S}$ value.}
	\label{FigHornsRevOBS}
\end{figure}

Hereafter, we compare the wind-farm models performance with SCADA data taken from Barthelmie et al\cite{Barthelmie2009}. These data consist in turbine power outputs averaged over three wind sectors (i.e, $\pm$ 5°, $\pm$ 10° and $\pm$ 15° ) centered on three mean wind directions (i.e, 221°, 270° and 312°). The power data are available only for turbines labelled in orange in Fig. \ref{FigHornsRevLES}(a). Fig.~\ref{FigHornsRevPowerRow} illustrates the modelled and observed turbine power ratios for the three transects. Overall, the observations show lower turbine power output when the turbine spacing is minimum, which occurs at 270°. Moreover, due to the limited number of turbines per row, the power ratios do not reach an asymptotic behaviour far downstream in the transect. Instead, they monotonically decrease. For all wind directions and wind sectors, the highest error in power estimates usually occurs for the second turbine of the row. On the other hand, for wind sectors of $\pm$10° and $\pm$15°, the models reproduce power losses quite well far downstream in the farm. To better illustrate the models performance, Fig. \ref{FigHornsRevOBS} displays the distribution of the difference in turbine power output between models predictions and SCADA data ($P_\mathrm{SCADA}$) shown in Fig.~\ref{FigHornsRevPowerRow}, which is computed as
\begin{equation}
\Delta_\mathrm{M\text{--}S} = \frac{P_\mathrm{Mod}^{\mathrm{t}} - P_\mathrm{SCADA}}{P_1}
\label{EqepsilonMS}
\end{equation}
where $P_\mathrm{Mod}^{\mathrm{t}}$ denotes the modelled turbine power, calculated using 1° increment, and further averaged over the studied sector. For instance, Fig. \ref{FigHornsRevOBS}(a) displays the distribution of $\Delta_\mathrm{M\text{--}S}$ encompassing the wind directions 221° $\pm$5°, 270° $\pm$5° and 312° $\pm$5°. Since $P_1$= 0.6 MW, a $\Delta_\mathrm{M\text{--}S}$ of 10\% corresponds to a difference in turbine power output between model and SCADA data of 0.06 MW (3\% of the turbine rated power). Note that $\Delta_\mathrm{M\text{--}S}$ is always zero for the most upwind turbine of the row, hence this value is not included in the statistics. All wind-farm models underestimate the turbine power if a narrow wind sector is considered (see Fig. \ref{FigHornsRevOBS}(a)). A similar behaviour was already observed in Fig. \ref{FigHornsRevLES}(b). However, if wider wind sectors are used, the models predictions are more accurate, with biases closer to zero and narrower interquartile ranges (see Fig. \ref{FigHornsRevOBS}(b,c)). These tendencies were also noticed by Barthelmie et al\cite{Barthelmie2009}. Overall, the linear superposition of velocity deficits and the new wake-merging method perform similarly. The most accurate models are again the Gaussian and double-Gaussian ones, which show error distributions with close to zero biases and small interquartile ranges in all cases. In fact, the very similar wake profile in the far-wake region obtained with these two models (see Fig.~\ref{FigSingleWakeModels}(b) and \ref{FigSingleWakeModels}(d)) makes them almost undistinguishable in terms of power outputs when $s_x>$ 4. Finally, both the Jensen and the super-Gaussian wake model show a strong negative bias, with $\Delta_\mathrm{M\text{--}S}$ values down to --30\%. 

\subsection{London Array} \label{LondonArray}
\begin{figure}[t!]
	\centering
	\begin{minipage}{\textwidth}
		\centering
		\includegraphics[width=0.26\textwidth]{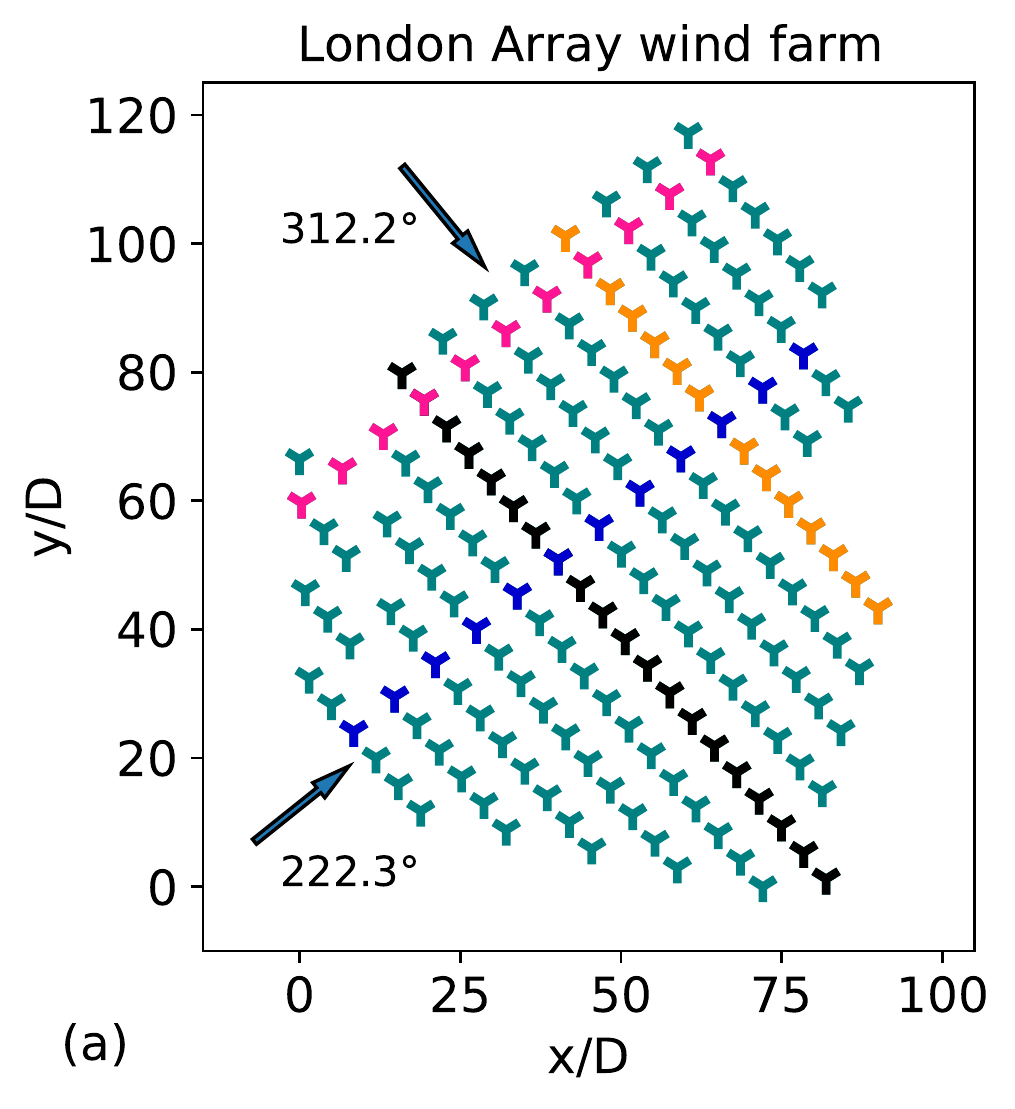}%
		\hspace{0.8\textwidth}
		\includegraphics[width=0.485\textwidth]{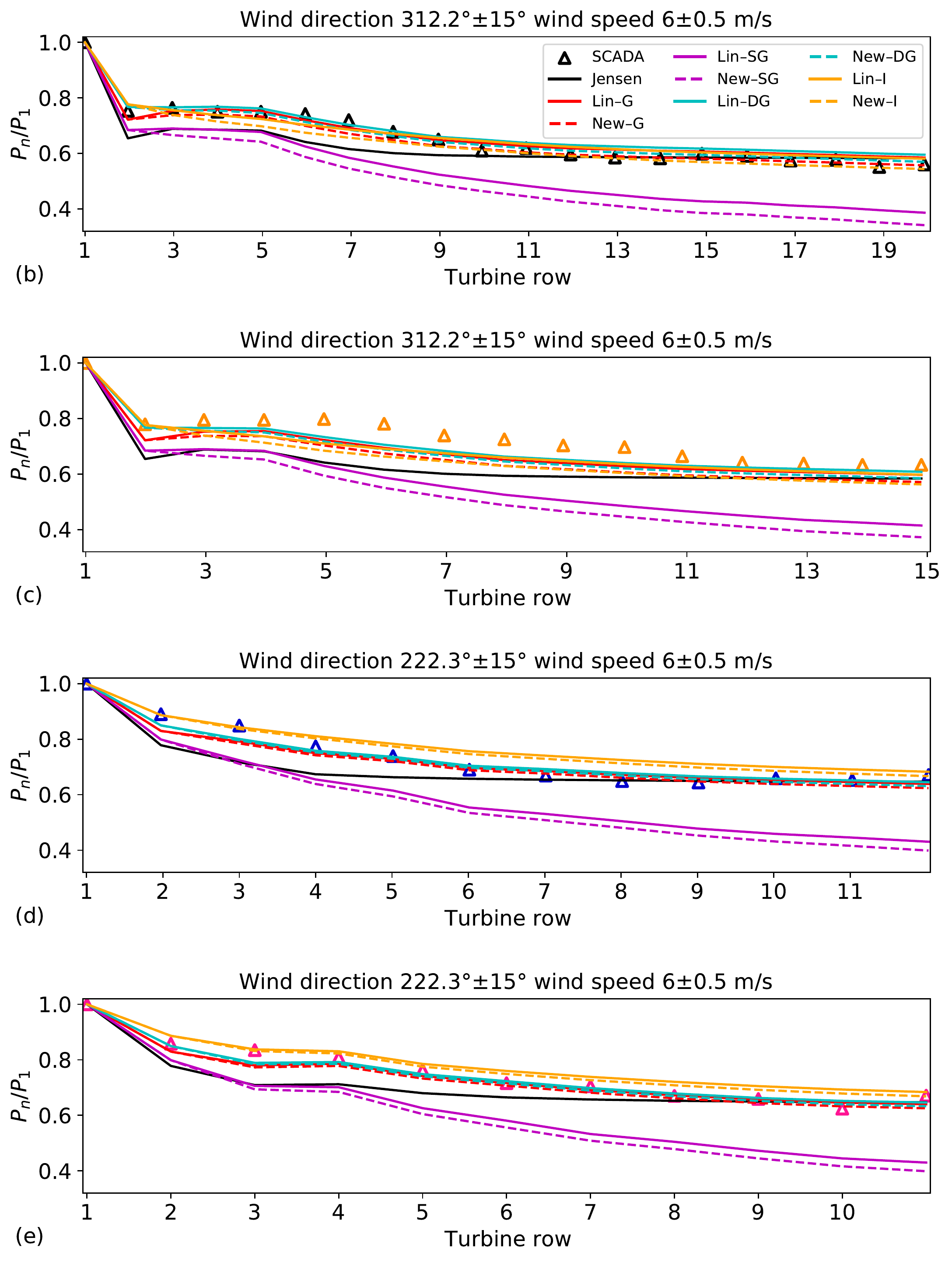}
		\hspace{0.015\textwidth}	
		\includegraphics[width=0.485\textwidth]{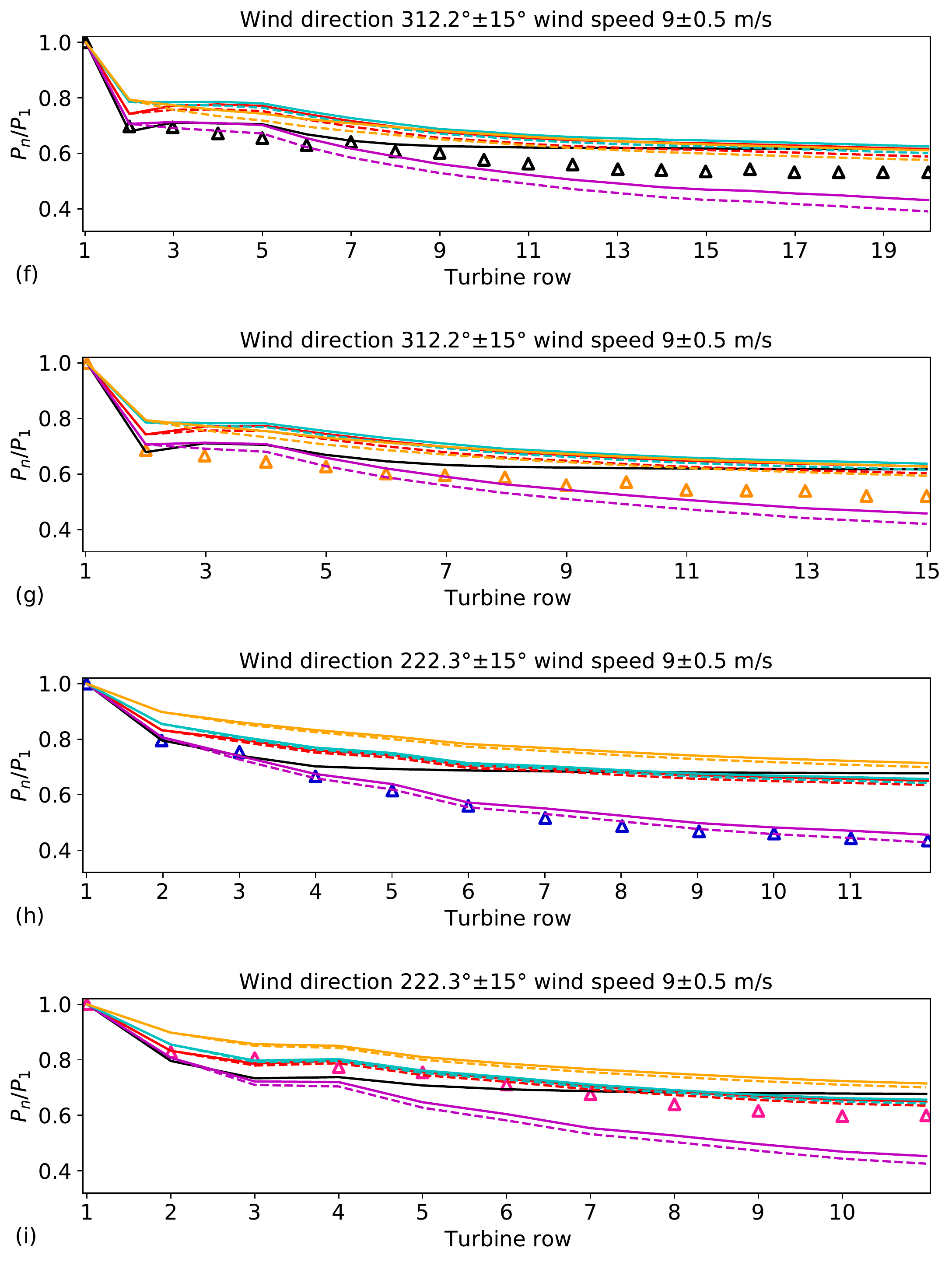}%
		\caption{(a) Layout of the London Array farm with distances normalized with the rotor diameter. The transects for which SCADA data are available are marked in black and orange for the north-western transects (312.2°) and in blue and pink for the south-western ones (222.3°). (b-e) Power ratios along the four transects for a wind speed of 6$\pm$0.5 m/s and (f-i) 9$\pm$0.5 m/s.}
		\label{FigLondonArrayOBS}
	\end{minipage}
\end{figure}

The London Array, with a capacity of 630 MW, is located 20 km off the Kent coast in the United Kingdom. The farm characteristics are listed in Table \ref{Table2} while the farm layout is displayed in Fig. \ref{FigLondonArrayOBS}(a). The SCADA data are available in Nygaard\cite{Nygaard2014} and consist in turbine power outputs measured at two different wind speeds (6 and 9 m/s) for four different transects (the ones in black, orange, blue and pink in Fig. \ref{FigLondonArrayOBS}(a)). The ambient turbulence intensity is measured by a met mast located upstream of the farm. The observed values are reported in Nygaard\cite{Nygaard2014}. The turbine spacings for the north- and south-western wind directions are 5.4 and 8.3 times the rotor diameter, respectively.

Figure \ref{FigLondonArrayOBS}(b-e) displays the modelled and observed turbine power ratios for the four transects at a wind speed of 6$\pm$0.5 m/s (i.e., we run the models for this range of wind speeds and we plot the averaged value). For all transects, the observations show a modest power drop between the first and the second turbine of the row. As usually happens in large farms, the power loss stabilises after approximately the tenth turbine of the row. Overall, the new wake-merging method predicts a slightly lower turbine power output than the linear superposition method. The Gaussian, double-Gaussian and Ishihara model perform similarly and show a very good agreement with observations. The Jensen model underestimates the power output of the firsts turbines in the row, but it captures the asymptotic behaviour in all transects. A better tuning of this model could possibly improve its performance (e.g., see Peña et al\cite{Pena2013}), but this is not in the scope of the current manuscript. As for the Horns Rev farm, the super-Gaussian wake model strongly underestimates the power output in all transects, being on average 30\% off from observations for the last turbine of the row.

The results obtained for a wind speed of 9$\pm$0.5 m/s are shown in Fig. \ref{FigLondonArrayOBS}(f-i). It is worth to notice that the models performance differ from the ones observed at 6$\pm$0.5 m/s, showing that it is bad practice to judge the quality of a model using data of a single row of turbines or using a unique wind speed value, as often done in literature. In fact, although some models show good agreement with observations for the firsts turbines in the row, they all fail in capturing the asymptotic behaviour reached far downstream in the transects. Fig. \ref{FigLondonArrayOBS}(h) shows that the observed power continues to drop through the array, in contrast to the previous cases. Nygaard\cite{Nygaard2014} excluded the possibility that the deep array effect (Barthelmie and Jensen\cite{Barthelmie2010b}, Schlez and Neubert\cite{Schlez2009}) could have caused such a drop and instead he claimed that this behaviour is due to the differences in inflow conditions present at 9 m/s between the north- and south-western transect. The super-Gaussian model fits extremely well the observations in such conditions, however it performs poorly in all other cases.

\begin{figure}[t!]
	\centering
	\begin{minipage}{\textwidth}
		\centering
		\includegraphics[width=0.35\textwidth]{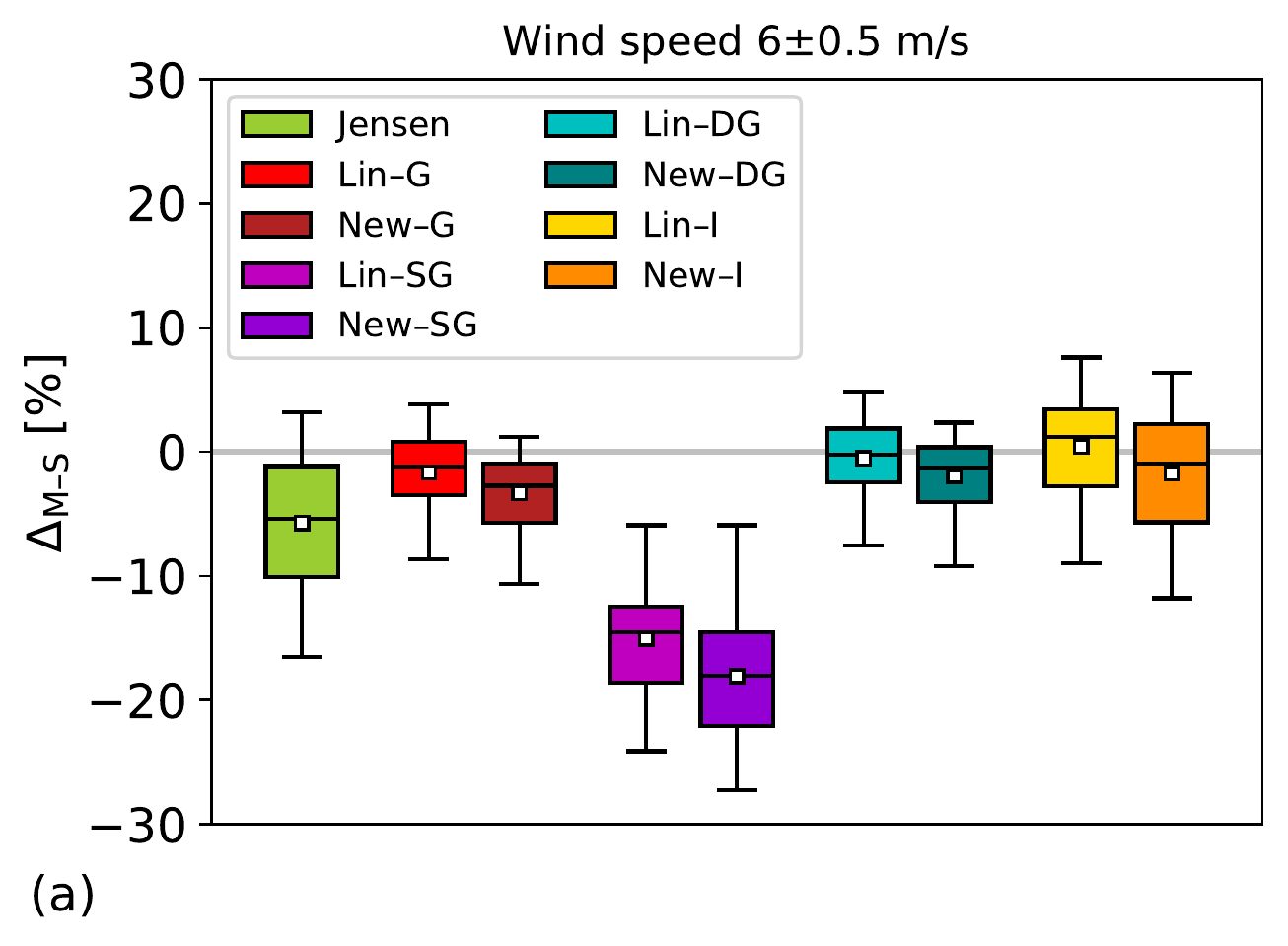}
		\hspace{0.1\textwidth}	
		\includegraphics[width=0.35\textwidth]{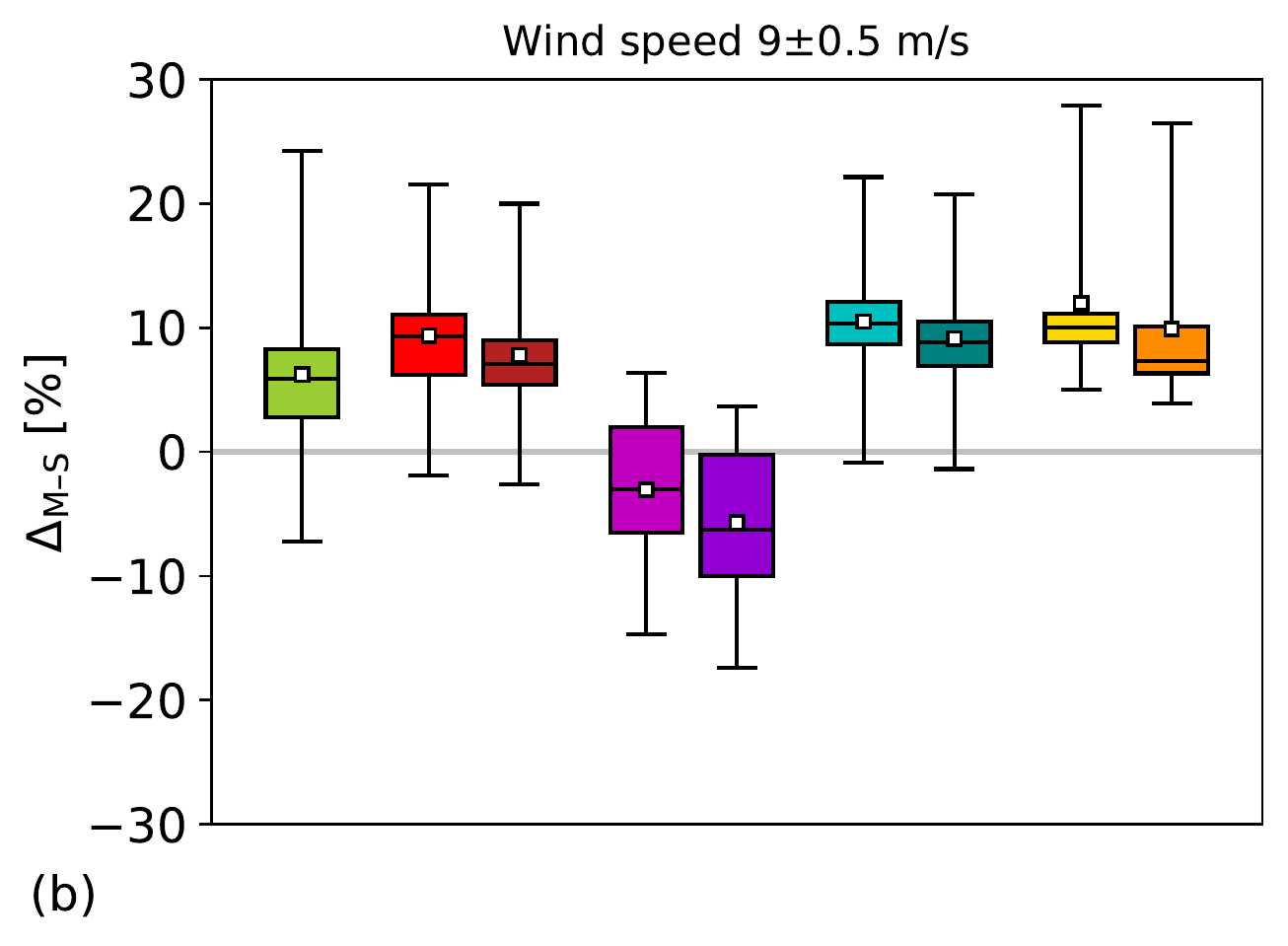}%
		\caption{Distribution of the difference in turbine power output between models predictions and SCADA data for a wind speed of 6$\pm$0.5 m/s (a) and 9$\pm$0.5 m/s (b). The difference in turbine power output is normalized with the freestream power. The box length represents the interquartile range. The horizontal black line and the white square denote the median and the mean, respectively. The whiskers include outliers, hence the caps represent the maximum and minimum $\Delta_\mathrm{M\text{--}S}$ value.}
		\label{FigLondonArrayOBSbox}
	\end{minipage}
\end{figure}

To better summarize the models performance, Fig. \ref{FigLondonArrayOBSbox} displays the distribution of the difference in turbine power output between models predictions and SCADA data shown in Fig.~\ref{FigLondonArrayOBS}. The freestream power $P_1$ is equal to 0.62 and 2.19 MW for the 6 and 9 m/s case, hence a $\Delta_\mathrm{M\text{--}S}$ of 10\% corresponds to a difference in turbine power output between model and SCADA data of 0.06 a and 0.22 MW, respectively (i.e., 1.6\% and 6.1\% of the turbine rated power). Since the error in power estimation for the most upwind turbine of the row is always zero, it is not included in the statistics. The box plots enlight the very similar performance between the two different wake-merging methods as well as the very accurate predictions obtained with the Gaussian, double-Gaussian and Ishihara wake model, particularly for 6 m/s wind speed. 

\subsection{Westermost Rough} \label{Westermost}
\begin{figure}[t!]
	\centering
	\begin{minipage}{\textwidth}
		\centering
		\includegraphics[width=0.26\textwidth]{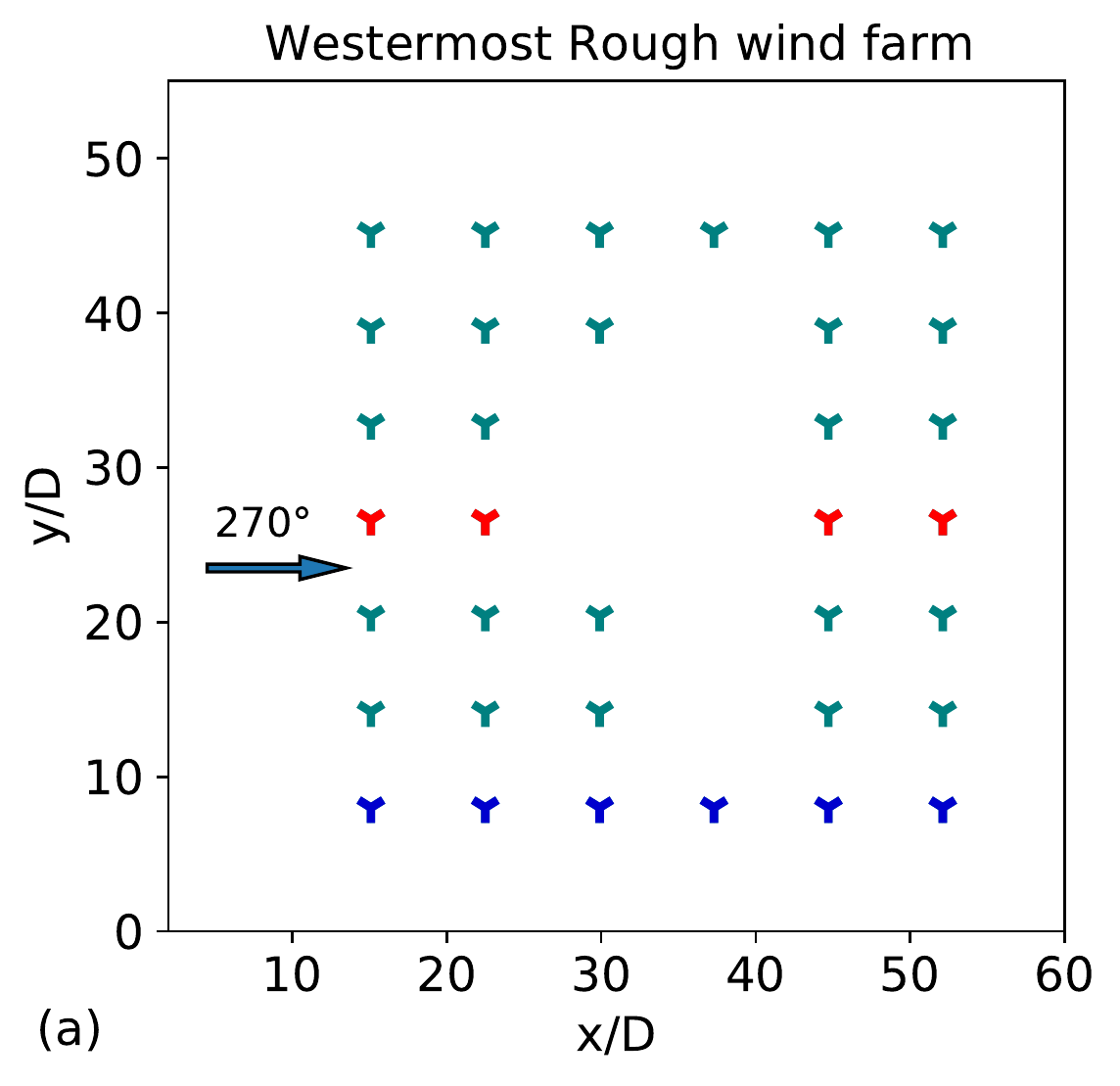}%
		\vspace{0.007\textwidth}
		\hspace{0.8\textwidth}
		\vspace{0.007\textwidth}
		\includegraphics[width=0.8\textwidth]{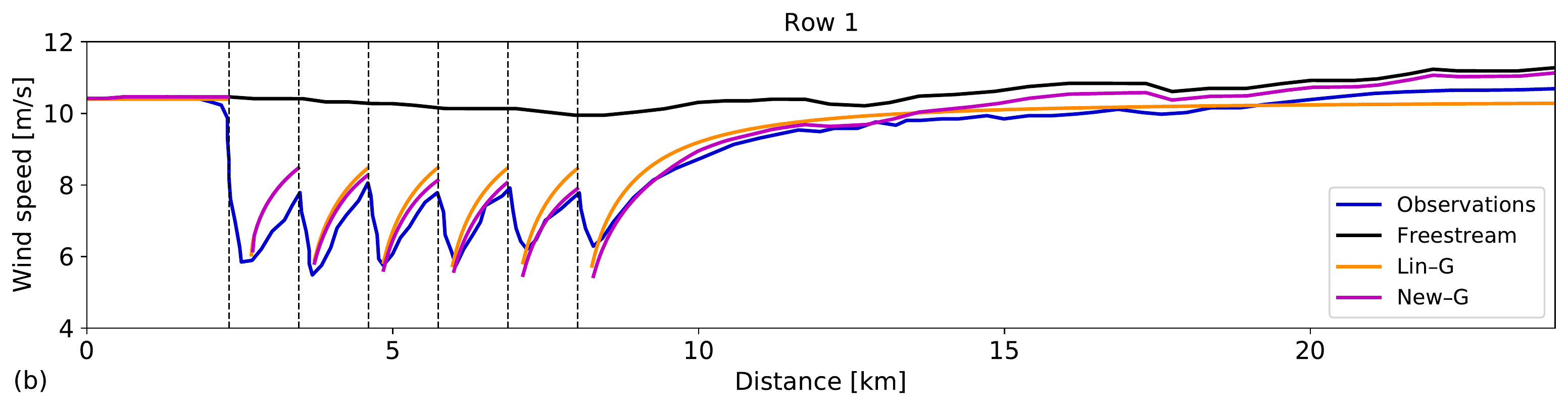}
		\hspace{0.028\textwidth}	
		\includegraphics[width=0.8\textwidth]{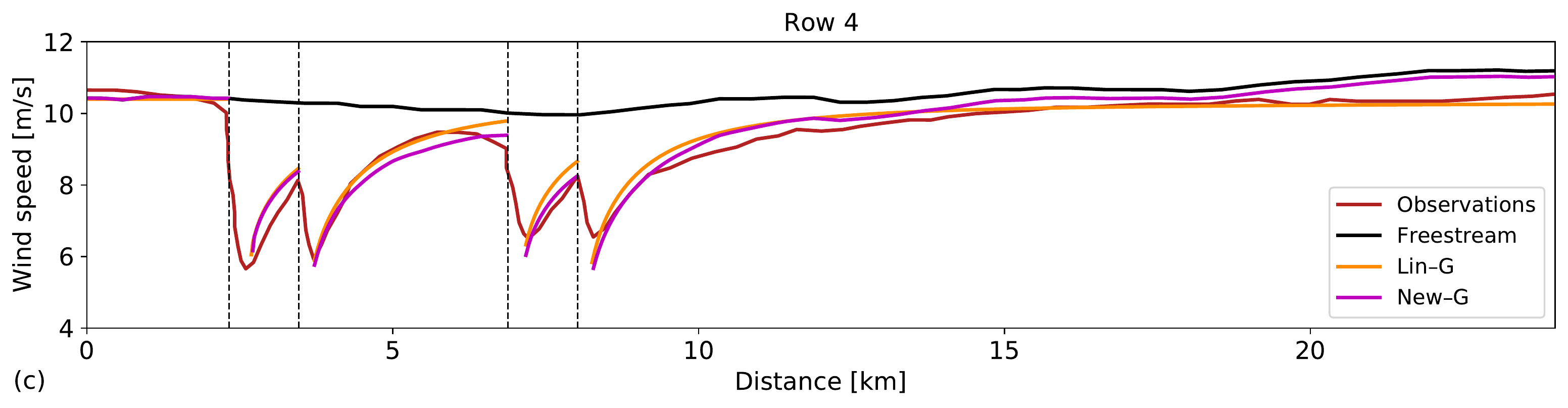}%
		\caption{(a) Layout of the Westermost Rough farm with distances normalized with the rotor diameter. The transects for which dual-Doppler data are compared with models predictions are marked in blue (row 1) and red (row 4). Freestream velocity, one-hour averaged dual-Doppler wind speed and velocity predicted with the new (New–G) and the linear (Lin–G) wake-merging method coupled with the Gaussian wake model for the blue (b) and red (c) transect. The vertical dashed lines indicate the turbine positions.}
		\label{FigWestermost1D}
	\end{minipage}
\end{figure}

The Westermost Rough farm, located 8 km off the Holderness coast in the United Kingdom, became operational in May 2015. The farm employs one of the largest offshore turbines to date with rotor diameter $D$= 154 m and hub height $z_h$= 106 m. The farm layout and characteristics are shown in Fig. \ref{FigWestermost1D}(a) and Table \ref{Table2}, respectively. In addition to the comparison between modelled and observed power outputs, we will asses the quality of the models also by analyzing the velocity fields through the farm and in the farm wake. The observation data are provided by Nygaard and Newcombe\cite{Nygaard2018} and consist in dual-Doppler radar wind speed at a horizontal plane at hub height. The data are averaged over a time window of 1 hour. The velocity field is reconstructed from 3 to 32 km from the shore (streamwise direction) and for 20 km in a direction parallel to the shore (spanwise direction). Therefore, the entire flow field in and around the farm (which is approximately 7$\times$7 km\textsuperscript{2}) is available. However, we will compare the models predictions only with two transects of the dual-Doppler data which span from 2 km upstream to 16 km downstream of the farm and run across the first and fourth row of turbines. Two additional transects run parallel and on both side of the farm at a distance of 10$D$ from the first and last row of turbines. Since these two transects do not cross the farm (and its wake), the velocity measured here is referred as freestream velocity. To derive the value of $U_\mathrm{b}$ at the first and fourth row of turbines, Nygaard and Newcombe\cite{Nygaard2018} used an interpolation technique. A plot of the dual-Doppler radar wind speed at hub height and more details about the measurements campaign are reported in Nygaard and Newcombe\cite{Nygaard2018}. Note that the authors do not provide the ambient turbulence intensity value, hence we assume $\mathrm{TI}_\mathrm{b}=$ 6\% which is in line with the values reported by Nygaard\cite{Nygaard2014} and Nygaard et al\cite{Nygaard2020}.

Figure \ref{FigWestermost1D}(b,c) shows the freestream velocity and the dual-Doppler wind speed measurements along transects which run through the first and fourth row of the farm together with the models predictions. To not overcomplicate the plot, we show the streamwise velocity of only two wind-farm models, i.e. linear superposition and new-wake merging method coupled with the Gaussian wake model (Lin–G and New–G). Note that the models streamwise velocity are averaged over an imaginary rotor with the same diameter of the turbines disk. The freestream velocity takes a quasi-constant value of 10.4 m/s in the farm induction region and it slightly decreases down to 10 m/s at the end of the farm. However, far downstream, the freestream velocity gradually increases up to 11.3 m/s. Since the farm is located only 8 km from the shore, the increase in freestream velocity is attributed to a coastal gradient induced by the difference in roughness between land and sea surface. Note that the freestream velocity (interpreted as background velocity in our study) only varies with approximately 1 m/s. The dual-Doppler wind speed along the blue and red transect show the typical sawtooth behaviour with maxima and minima located upstream and downstream of the turbines. The velocity reduction in front of the third turbine of row 4 is attributed to axial induction. The same flow behaviour is also observed for the first turbine of row 1 and 4. The linear superposition of velocity deficits coupled with the Gaussian wake model (Lin–G) does not account for spatially varying background velocity. Instead, it relies on a global wind speed value measured several hundreds of meters upstream of the farm, which we fix to 10.4 m/s. On the other hand, the same single-wake model coupled with the new wake-merging method (New–G) uses the spatially varying freestream velocity as input. The two wind-farm models predictions begin to diverge from the second turbine of the row. The linear wake-merging method overestimates the velocity at turbine locations in both row 1 and 4. Contrarily, the new superposition method shows very good agreement with observations. Also, the wind speed far downstream the farm converges to the freestream velocity and not to the upstream wind speed value, which is a more realistic representation of the reality. Note that the Gaussian wake model does not conserve momentum in the near-wake region. For this reason, the wind speed models predictions are plotted only in the far-wake, which corresponds to the region of interest of our study. Also, the model does not account for axial induction. This is an important topic for further research.  

\begin{figure}[t!]
	\centering
	\includegraphics[width=0.95\textwidth]{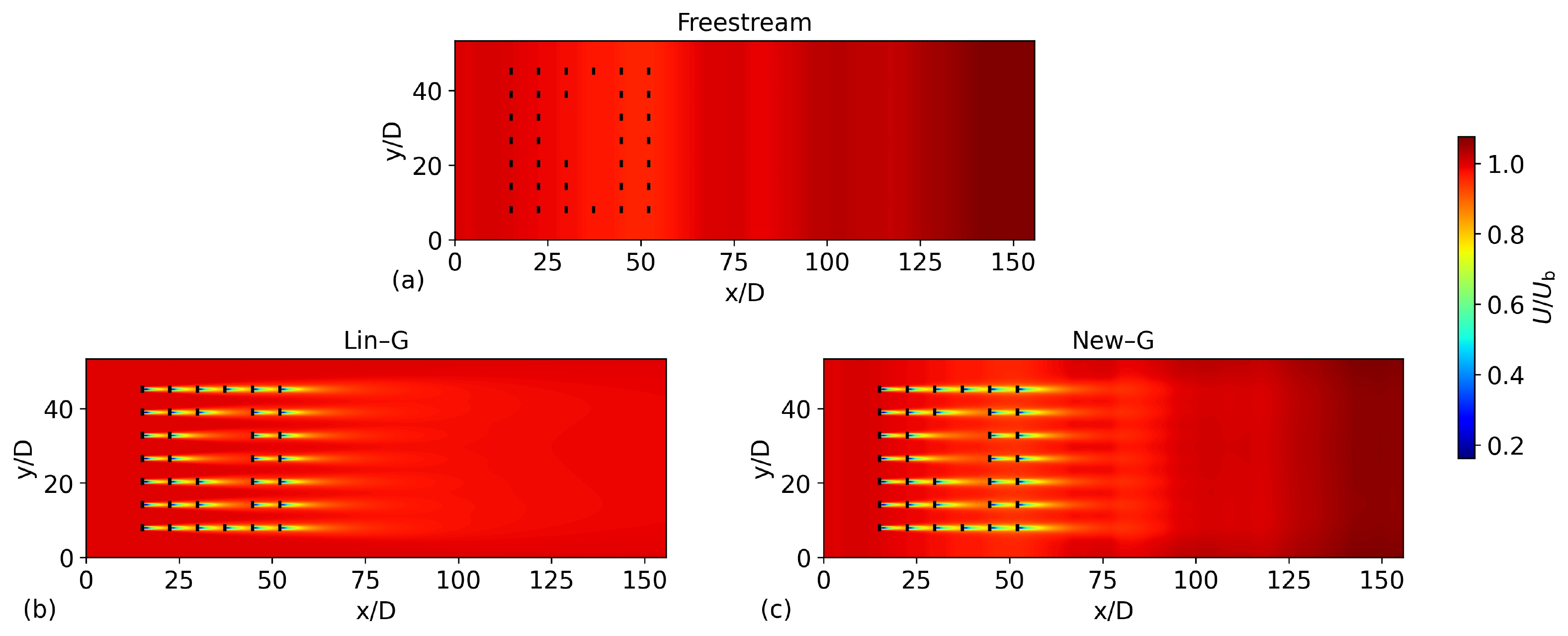}%
	\caption{(a) Freestream wind speed and (b,c) velocity field predicted with the linear and new wake-merging method coupled with the Gaussian wake model, respectively. The velocity is normalized with the wind speed measured 2$D$ upstream of the farm. The black lines denote the wind-turbine rotor locations.}
	\label{FigWestermost2D}
\end{figure}

A two-dimensional plot displaying the freestream velocity at hub height is shown in Fig. \ref{FigWestermost2D}(a), where the increase in wind speed along the streamwise direction due to coastal gradient effects is noticeable. Note that we plot the value of $U_\mathrm{b}$ interpolated along the transect which crosses the fourth row of turbines and we assume it constant along the spanwise direction. Moreover, the velocity field is normalized with the wind speed measured 2$D$ upstream of the farm. Fig. \ref{FigWestermost2D}(b,c) illustrates the velocity at a horizontal plane at hub height obtained with the Lin–G and New–G models, respectively. Since both wake-merging methods are coupled with the same single-wake model, the observed differences only reside in how the methods overlap preceding wakes and deal with the freestream velocity.

Finally, we define 
\begin{equation}
\Delta_\mathrm{M\text{--}O} = \frac{P_\mathrm{Mod}^\mathrm{t} - P_\mathrm{Obs}}{P_1}
\label{EqepsilonMO}
\end{equation}
where $P_\mathrm{Mod}^\mathrm{t}$ and $P_\mathrm{Obs}$ refer to modelled and observed single-turbine power outputs. A $\Delta_\mathrm{M\text{--}O}$ of 10\% corresponds to a difference between modelled and observed turbine power of 0.5 MW, since $P_1$=5.1 MW. Note that SCADA data are not available for this case study. Therefore, to obtain $P_\mathrm{Obs}$, we use the turbine power curve where the velocity input refers to the turbine inflow velocity (i.e., the observed velocity in correspondence of the vertical dashed line in Fig. \ref{FigWestermost1D}(b,c)). For completeness, we have included all wind-farm models in this analysis. Overall, the new wake-merging method outperforms the linear superposition one. In fact, since the freestream velocity decreases over the wind-farm area (see Fig. \ref{FigWestermost1D}(b,c)), the new model predicts lower turbine power outputs, showing biases closer to zero than the ones obtained with the linear method, which are in average 10\% off from observations. As in the previous analysis, the Lin–SG model strongly underestimates the turbine power outputs. Consequently, in light of the above reasoning, the New–SG model shows an even higher bias. However, we believe that a better tuning of the wake model would change this result, but this is out of the scope of the current manuscript. The better agreement with observations achieved with the new wake-merging method (except when coupled with the super-Gaussian wake model) once again points out the importance of accounting for a spatially varying background velocity. The power output predictions would further improve if axial induction effects would be included in the analytical models.

\begin{figure}[t!]
	\centering
	\includegraphics[width=0.42\textwidth]{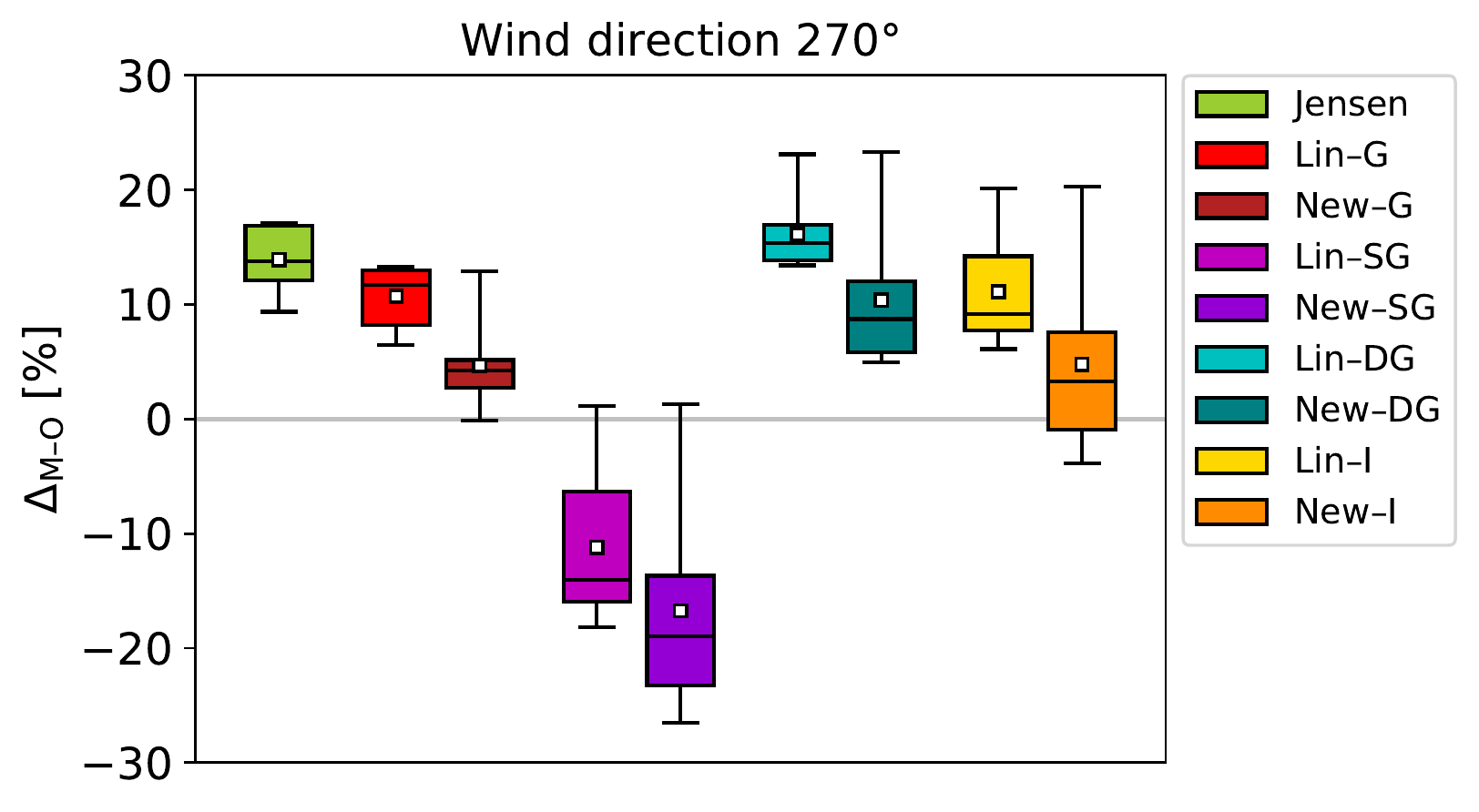}%
	\caption{Distribution of the difference in turbine power output between models predictions and dual-Doppler data. The difference in turbine power output is normalized with the freestream power. The box length represents the interquartile range. The horizontal black line and the white square denote the median and the mean, respectively. The whiskers include outliers, hence the caps represent the maximum and minimum $\Delta_\mathrm{M\text{--}O}$ value.}
	\label{FigWestermostPower}
\end{figure}

\section{Conclusions}\label{conclusion}
In the current study, we proposed a new wake-merging method for predicting wake losses and farm power outputs. The waked flow through a farm is expressed as a function of a spatially varying background velocity. Hence, the new superposition method does not rely on a global wind-speed value (which is usually taken upstream of the farm). Instead, it superimposes the wake velocity deficits generated by turbines on a heterogeneous velocity field. This is of fundamental importance for achieving better farm power predictions and consequently more accurate annual energy yield assessments. In fact, the meso-scale velocity gradients that develop in coastal regions or in proximity of orographic obstacles make wind farms operating in velocity fields which are rarely uniform. The proposed superposition method is consistent with self-similarity in the wake of every turbine in the farm and is momentum-conserving under the assumptions of slow varying background velocity and large enough turbine spacings.

LES data, Dual-Doppler radar measurements and SCADA data from the Horns Rev, London array and Westermost Rough farm were used for validating the new wake-merging method. To this end, the new and the linear\cite{Niayifar2016} superposition method were coupled with the Gaussian \cite{Bastankhan2014}, super-Gaussian \cite{Blondel2020}, double-Gaussian \cite{Schreiber2020} and Ishihara \cite{Ishihara2018} single-wake model. The Jensen model coupled with quadratic\cite{Voutsinas1990} superposition was also included in the analysis, as an additional point of reference. A homogeneous background velocity field was used in the Horns Rev and London Array case study. In such conditions, the new wake-merging method performed similarly to the linear one. The most accurate power predictions were provided by the Gaussian and double-Gaussian wake model, closely followed by the Ishihara model. The Jensen wake model showed a negative bias in power estimates for the majority of wind speeds and wind directions here analyzed. The model predictions may be improved by expressing the wake expansion coefficient as a function of the turbulence intensity. The super-Gaussian wake model was the one that diverged the most from observations, underestimating the power output in all circumstances. We believe that a better tuning of the model coefficients could improve its performance. Finally, a heterogeneous background velocity field was used in the Westermost Rough case study. The new wake-merging method outperformed the linear one in such conditions, showing very good agreement with dual-Doppler radar wind speed and confirming the importance of accounting for a spatially varying background velocity when looking at farm power assessments.

Future research will focus on a more extensive validation of the proposed wake-merging method in presence of a heterogeneous background velocity field. The Anholt wind farm is a suitable candidate for this type of study due to the strong coastal gradient present in its location \cite{Pena2018}. Also, the new wake-merging model could account for velocity gradients generated by self-induced gravity waves. Hence, we plan to couple the new wind-farm model with the recently developed mid-fidelity three-layer model\cite{Allaerts2019} in the future. To further improve the performance of wind-farm flow models, a more general parametrization of the wake growth rate is necessary. Moreover, the inclusion of axial induction effects would improve the models reliability and representation of reality in the vicinity of the turbines.

\section*{Acknowledgments}
The authors acknowledge support from the Research Foundation Flanders (FWO, grant no. G0B1518N), and from the project FREEWIND, funded by the Energy Transition Fund of the Belgian Federal Public Service for Economy, SMEs, and Energy (FOD Economie, K.M.O., Middenstand en Energie).

\subsection*{Author contributions}
L.L. and J.M. jointly developed the new wake-merging method and set up the simulation studies. L.L. performed code implementations and carried out the simulations. L.L. and J.M. jointly wrote the manuscript.

\subsection*{Conflict of interest}
The authors declare that they have no conflict of interest.

\appendix

\section{Comparison between linear superposition of velocity deficits and new wake-merging method}\label{AppNewNia} 
The aim of this section is to compare the velocity field simulated with two different wake-merging methods, namely the one proposed by Niayifar and Porté-Agel\cite{Niayifar2016} and the one derived in this manuscript. The linear superposition method does not support varying background velocity field. Therefore, to conduct a fair comparison, we use a uni-directional homogeneous $U_\mathrm{b}$ and we set it to 10 m/s. The turbine rotor diameter is $D$= 154 m and the turbine thrust set-point is $C_\mathrm{T}$=0.85. The ambient turbulence intensity is fixed to $\mathrm{TI}_\mathrm{b}$= 12\% and the added turbulence through the farm is evaluated with the model proposed by Niayifar and Porté-Agel\cite{Niayifar2016}. Moreover, both wake-merging methods adopt a Gaussian shape function.

\begin{figure}[t!]
	\centering
	\includegraphics[width=0.95\textwidth]{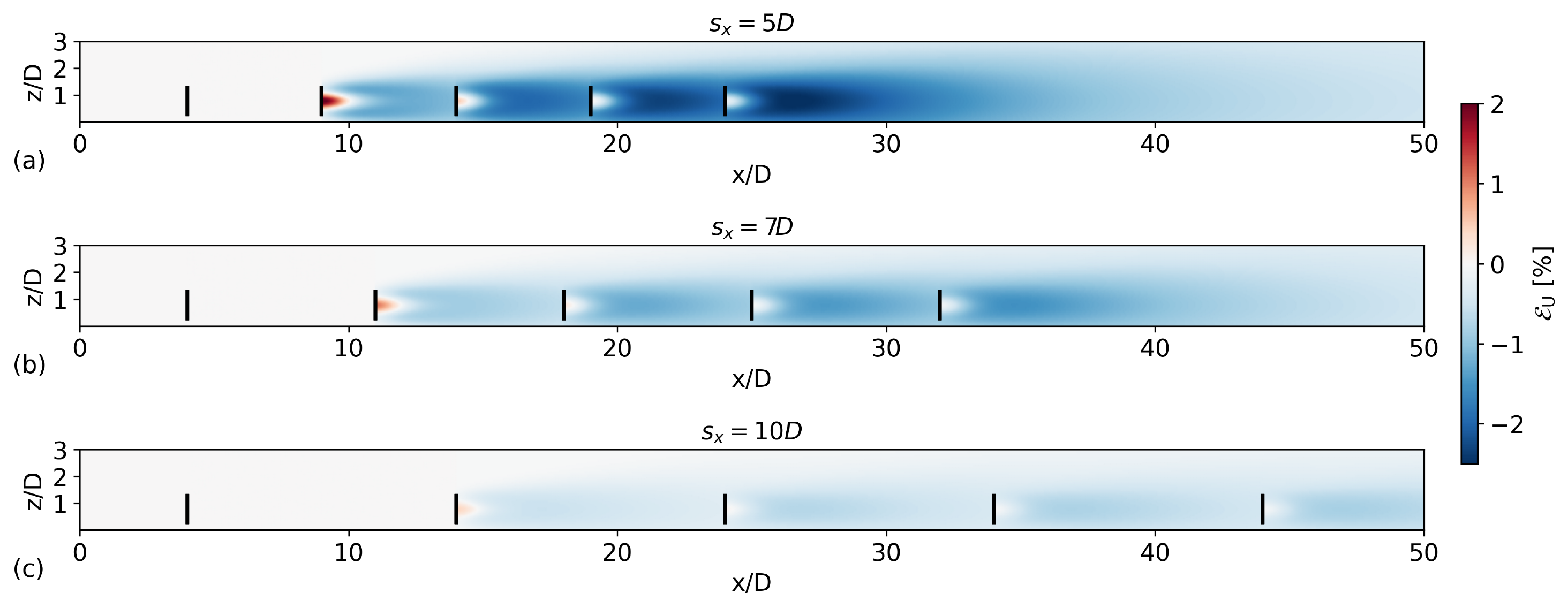}%
	\caption{Difference between velocity fields computed with the new and the linear superposition method coupled with the Gaussian wake model. The difference is visualized at a vertical plane normal to the wind turbine at zero span for a turbine spacing $s_x$ of (a) 5$D$, (b) 7$D$ and (c) 10$D$. The black lines denote the wind-turbine rotor locations.}
	\label{FigAppendix1}
\end{figure}

The difference between velocity fields is computed as
\begin{equation}
\mathcal{E}_\mathrm{U}(\boldsymbol{x}) = \frac{U^{\mathrm{New}}(\boldsymbol{x}) - U^{\mathrm{Lin}}(\boldsymbol{x})}{U_\mathrm{b}}
\end{equation}
where $U^{\mathrm{New}}(\boldsymbol{x})$ and $U^{\mathrm{Lin}}(\boldsymbol{x})$ denote the velocity through the farm predicted by the new and the linear superposition method, respectively. Fig.~\ref{FigAppendix1} displays $\mathcal{E}_\mathrm{U}(\boldsymbol{x})$ at a vertical plane normal to the wind turbine at zero span for a row of 5 turbines. Since the two wake-merging methods are analytically equal for a single turbine, the velocity fields are identical up to the second turbine of the row. However, even though a uni-directional homogeneous background velocity is used, the models clearly differ far downstream in the row. Overall, the new superposition method predicts a smaller velocity through the farm specially for a small streamwise turbine spacing, as displayed in Fig.~\ref{FigAppendix1}(a). Note that $\mathcal{E}_\mathrm{U}(\boldsymbol{x})$ assumes positive values only in the near-wake region of the second turbine of the row. However, the thrust coefficients are written as an error function of the streamwise coordinate in this region \cite{Zong2020}, since the Gaussian wake model is not momentum-conservative for $C_\mathrm{T} > D^2/8\delta^2$ (which usually happens when $x/D<1.5$). Therefore, the models predictions are not relevant in this region. On the other hand, Fig.~\ref{FigAppendix1}(c) shows that the models predictions become comparable when turbines are widely spaced ($\mathcal{E}_\mathrm{U} \approx $--$0.2\%$). In fact, the wake function converges to zero for large streamwise distances, redimensioning the differences between superposition methods. Note that $\mathcal{E}_\mathrm{U}(\boldsymbol{x})$ shows an axisymmetric distribution for all turbine spacings since we have considered no vertical shear and we have neglected ground-wake interactions.

\begin{figure}[t!]
	\centering
	\includegraphics[width=0.9\textwidth]{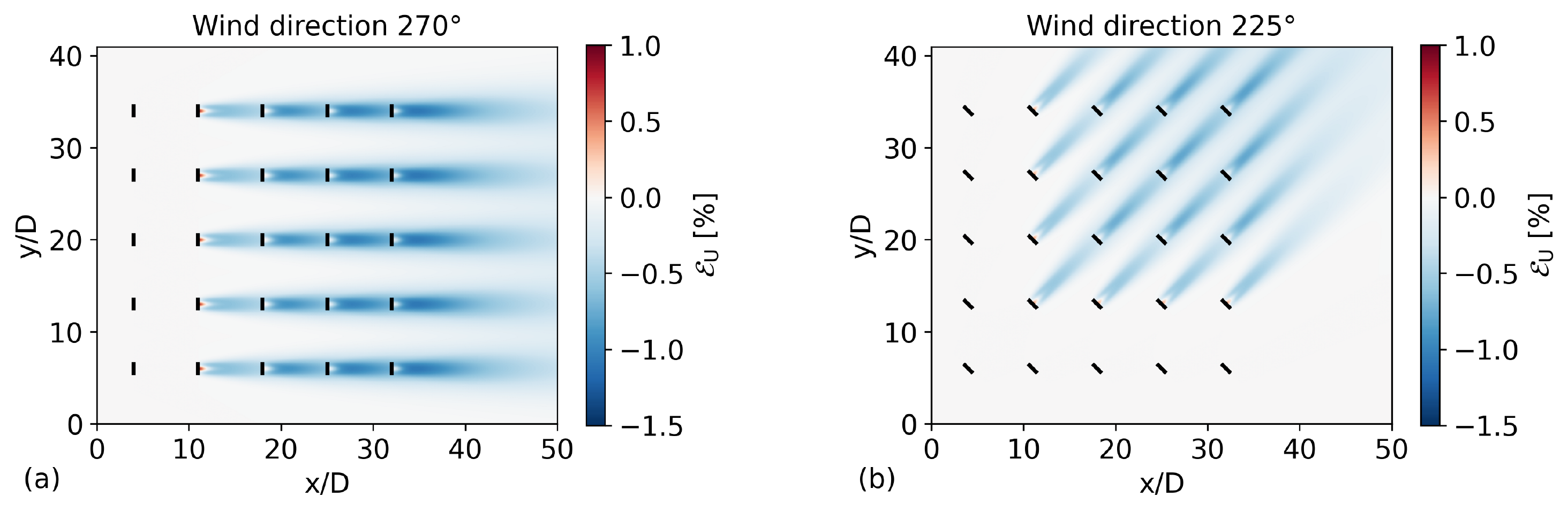}%
	\caption{Difference between velocity fields computed with the new and the linear superposition method coupled with the Gaussian wake model. The difference is visualized at a horizontal plane at hub height for a wind direction of (a) 270° and (b) 225°. The black lines denote the wind-turbine rotor locations.}
	\label{FigAppendix2}
\end{figure}

Figure~\ref{FigAppendix2} illustrates the distribution of $\mathcal{E}_\mathrm{U}(\boldsymbol{x})$ at a horizontal plane at hub height for a farm of 25 turbines with streamwise and spanwise spacing of $s_x = s_y=$ 7. The models setup and atmospheric conditions correspond to the ones detailed above. The new superposition method predicts smaller velocity than the linear one far downstream in the farm, specially through the centerline of the wakes where $\mathcal{E}_\mathrm{U} \approx$ --1.5\%. As mentioned above, there is no difference between the two models up to the second row of turbines. Moreover, $\mathcal{E}_\mathrm{U}(\boldsymbol{x})$ assumes lower values for a wind direction of 225° since the turbine spacing is larger than the one at  270°.

\newpage

\end{document}